\documentclass[twocolumn,aps,pra,physrev,groupedaddress,superscriptaddress,amsmath,amssymb,showpacs,noeprint]{revtex4-2}

\usepackage{graphicx}
\usepackage{dcolumn}
\usepackage{bm}
\usepackage[dvipsnames]{xcolor}
\usepackage{hyperref}
\hypersetup{colorlinks,linkcolor={MidnightBlue},citecolor={MidnightBlue},urlcolor={MidnightBlue}}  

\usepackage{physics}
\usepackage{amsfonts}
\usepackage{verbatim}
\usepackage{amsmath}
\usepackage{csquotes}
\usepackage{color}
\usepackage{soul}
\usepackage{amsthm}
\usepackage{bm}
\usepackage{graphicx}
\usepackage[normalem]{ulem}

\newcommand{\prt}[1]{\left(#1\right)}

\newcommand{\prtq}[1]{\left[#1\right]}

\newenvironment{equations}
{\begin{equation}\begin{aligned}}
{\end{aligned}\end{equation}\ignorespacesafterend}

\newcommand{\nh}{n_h}
\newcommand{\nc}{n_c}
\newcommand{\nr}{n_r}
\newcommand{\ns}{n_m}

\newcommand{\uh}{u_h}
\newcommand{\uc}{u_c}

\newcommand{\ar}{a_r}
\newcommand{\ard}{a_r^\dagger}

\newcommand{\am}{a_m}
\newcommand{\amd}{a_m^\dagger}

\newcommand{\ac}{a_c}
\newcommand{\acd}{a_c^\dagger}
\newcommand{\ah}{a_h}
\newcommand{\ahd}{a_h^\dagger}

\newcommand{\thrm}{\textup{th}}

\newcommand{\Vm}{V_\textup{max}}
\newcommand{\VmedMax}{V_{m,\textup{max}}}

\newcommand{\mdelta}{{\delta_m}}

\begin{document}

\title{Power maximization of two-stroke quantum thermal machines}

\author{Nicol\`o Piccione}
\email{nicolo.piccione@univ-fcomte.fr}
\affiliation{Institut UTINAM, CNRS UMR 6213, Universit\'{e} Bourgogne Franche-Comt\'{e}, Observatoire des Sciences de l'Univers THETA, 41 bis avenue de l'Observatoire, F-25010 Besan\c{c}on, France}

\author{Gabriele De Chiara}
\affiliation{Centre  for  Theoretical  Atomic,  Molecular  and  Optical  Physics, Queen's  University  Belfast,  Belfast  BT7 1NN,  United  Kingdom}

\author{Bruno Bellomo}
\affiliation{Institut UTINAM, CNRS UMR 6213, Universit\'{e} Bourgogne Franche-Comt\'{e}, Observatoire des Sciences de l'Univers THETA, 41 bis avenue de l'Observatoire, F-25010 Besan\c{c}on, France}

\begin{abstract}
We present a detailed study of quantum thermal machines employing quantum systems as working substances.
In particular, we study two different types of two-stroke cycles where two collections of identical quantum systems with evenly spaced energy levels are initially prepared at thermal equilibrium by putting them in contact with a cold and a hot thermal bath, respectively.
The two cycles differ in the absence or the presence of a mediator system, while, in both cases, non-resonant exchange Hamiltonians are exploited as particle interactions.
We show that the efficiencies of these machines depend only on the energy gaps of the systems composing the collections and are equal to the efficiency of \enquote{equivalent} Otto cycles. Focusing on the cases of qubits or harmonic oscillators for both models, we maximize the engine power and analyze, in the model without the mediator, the role of the waiting time  between subsequent interactions.
It turns out that the case with the mediator can bring performance advantages when the interaction times are comparable with the waiting time of the correspondent cycle without the mediator.
We find that in both cycles, the power peaks of qubit systems can surpass the Curzon-Ahlborn efficiency.
Finally, we compare our cycle without the mediator with previous schemes of the quantum Otto cycle showing that high coupling is not required to achieve the same maximum power.
\end{abstract}

\maketitle

\section{Introduction}

The study of thermodynamic cycles at the quantum level is a current topic of interest~\cite{Vinjanampathy2016,BookBinder2018,Mitchison2019,deffner2019quantum} and various kinds of quantum thermal machines have been proposed in the last years \cite{Allahverdyan2010,Campisi2015,Leggio2015, Reid_2017,Cakmak2019,Abah2020,Molitor2020Stroboscopic}. In particular, much effort has been devoted to the tradeoff between power and efficiency, with power usually increased at the expense of efficiency. The efficiency of a thermal engine operating between two thermal baths at different temperatures is limited by the Carnot value $\eta_C = 1 - T_h/T_c$, where $T_h$ is the temperature of the hot bath and $T_c$ of the cold one ($T_h > T_c$). However, the Carnot limit is a very general bound usually attained for ideal machines with vanishing power~\cite{BookBinder2018}. For an engine operating at maximum power, the efficiency is not bounded by a universal value, and its maximum value is model dependent. Often, the efficiency at maximum power is limited by the so-called Curzon-Ahlborn efficiency~\footnote{Most papers in literature denote this efficiency as the Curzon-Ahlborn efficiency. Others name it the Chambadal-Novikov efficiency or Chambadal-Novikov-Curzon-Ahlborn efficiency. In this paper, we will use the more common name.} $\eta_{CA} = 1 - \sqrt{T_h/T_c}$~\cite{Curzon1975,VanDenBroeck2005,Esposito2009,Esposito2010,Wang2015,Shiraishi2016,Shiraishi2017}. However, this limit is not always valid~\cite{VanDenBroeck2005,Esposito2009,Esposito2010,Shiraishi2016,Shiraishi2017,BookBinder2018} and can be surpassed~\cite{Allahverdyan2010,Campisi2015,Erdman2019Maximum}.

When studying the power of a discrete quantum heat device, i.e., operating with finite-time strokes, the time allocated for thermalization typically plays an important role often representing the main contribution to the waiting time between subsequent strokes. In most cases, such a waiting time can be so high that all the other cycle steps can be considered instantaneous~\cite{Allahverdyan2010,Campisi2015,Uzdin_2016}. On the other hand, there is an ongoing effort to reduce the thermalization times, especially within the quantum Carnot cycle~\cite{Dann2019,Dann2020PRA,Dann2020NJP,Pancotti2019}, and to optimize the power output~\cite{Menczel2019}. In some cases, however, the waiting time is assumed negligible compared to other times, as done in other proposals of the quantum Otto cycle employing shortcuts to adiabaticity~\cite{Abah2017,Cakmak2019,Abah2020}, reducing the time necessary to vary the Hamiltonian of the working fluid.

In this paper, we propose two quantum thermal machines based on two-stroke cycles. In both models, two collections of identical quantum systems with constant and evenly spaced energy levels are initially prepared in equilibrium by bringing them in contact with a cold and a hot thermal bath, respectively. In the first scheme we propose, a system of one collection interacts with a system of the other one, and then they thermalize. The presence of many copies in each collection can change the magnitude of the effective waiting time with respect to the interaction time. In the second scheme, the two collections do not interact directly anymore but by means of a mediator system. The mediator interacts alternately with some systems from one collection and then with some from the other one. While the mediator interacts with one collection, the systems of the other collection can interact with their own thermal bath, thus amortizing the waiting time.

Under the general assumption that the various interactions preserve the number of excitations, we show that the efficiencies of the two cycles depend only on the Bohr frequencies of the two collections and are equal to the one of an \enquote{equivalent} Otto cycle. Remarkably, this is obtained without the need for time intervals where these  Bohr frequencies or energy gaps must be varied. With a specific choice of the interaction Hamiltonian, we investigate for both models the problem of power optimization, focusing on the case of qubits and harmonic oscillators. One of the novel aspects of our paper is a detailed analysis of the role of the waiting time between consecutive interactions, especially in relation to the maximum achievable power, an aspect which has often been overlooked in the literature. For the first model, we also compare our results with those of \enquote{equivalent} Otto cycles. Concerning the second model, instead, we show that the presence of a mediator can bring a performance advantage for a specific range of the waiting time in the correspondent cycle without the mediator. Moreover, a machine with the mediator system could be more easily implemented, depending on the experimental platform.

The paper is organized as follows. In Sec.~\ref{sec:ModelNoMediator}, we describe the model without the mediator, analyze its efficiency, and maximize its power. Then, in Sec.~\ref{sec:CycleMediator}, we perform a similar study for the model with the mediator and show that its presence can lead to performance advantages. In Sec.~\ref{sec:ComparisonOttoCycle}, we compare the cycle without the mediator with quantum Otto cycles which make use of shortcuts to adiabaticity. Finally, in Sec.~\ref{sec:Conclusions}, we briefly state the conclusions of our work, while we provide some details of our analysis and some calculations in several Appendices.

\section{\label{sec:ModelNoMediator}The model without mediator}

\begin{figure}[t]
	\centering
	\includegraphics[width=0.48\textwidth]{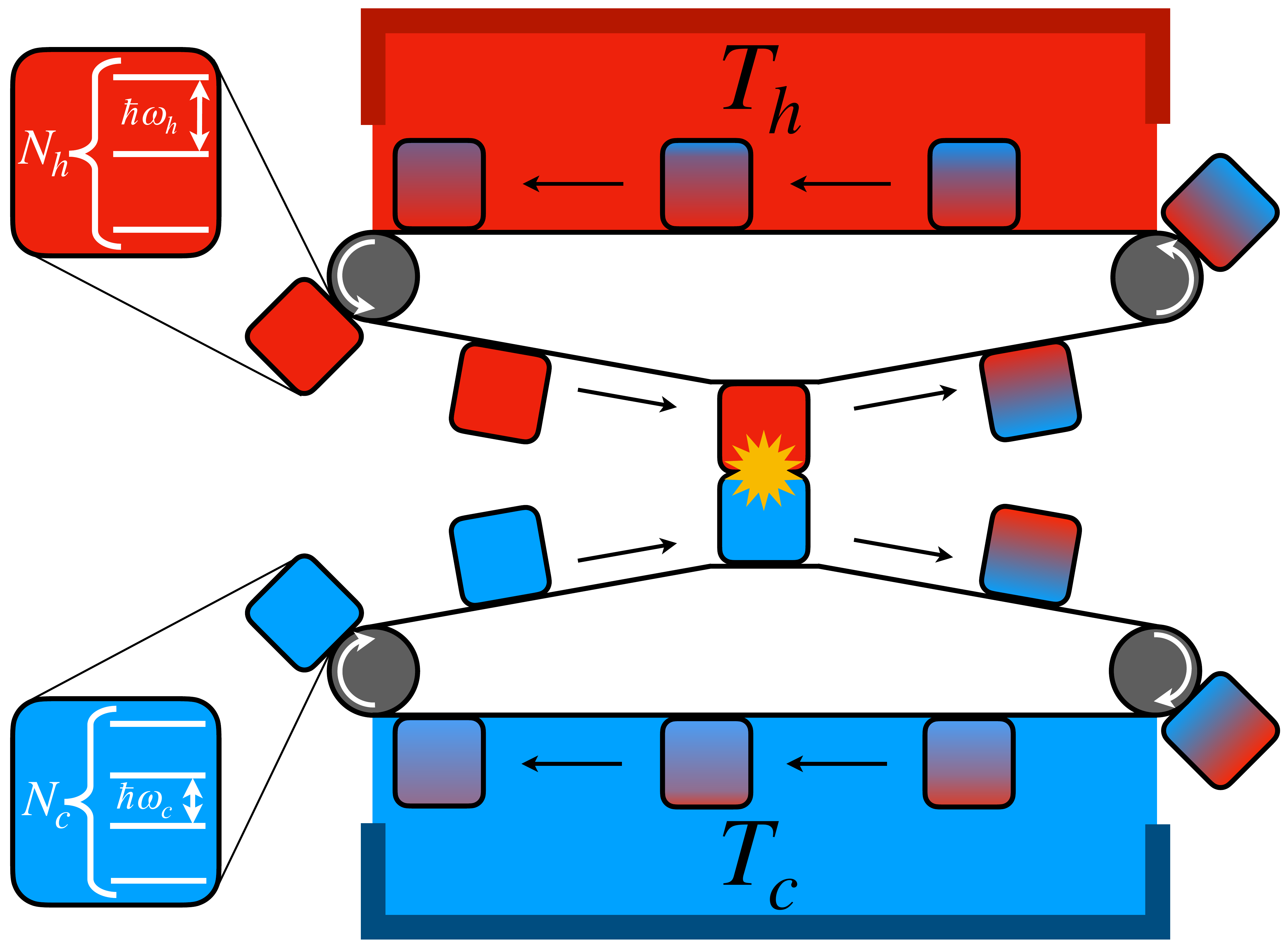}
	\caption{A scheme of the cycle with no mediator. Couples of systems belonging to each of the two collections interact for a time $\tau$ and then thermalize, sequentially. With enough couples, the thermalization of each one of them occurs during the collisions of other couples so that the waiting time $t_w$ can be greatly amortized and can become even negligible compared to the interaction time $\tau$.}
	\label{fig:No_Mediator}
\end{figure}

In this model, depicted in Fig.~\ref{fig:No_Mediator}, the thermal machine consists of two collections $C_c$ and $C_h$ of copies of, respectively, quantum systems $S_c$ and $S_h$, with  Hilbert space dimensions, respectively, $N_c$ and $N_h$. The systems of the two collections, when non-interacting, are in contact with a cold bath at temperature $T_c$ and a hot bath at temperature $T_h$, respectively. The systems $S_c$ and $S_h$ have evenly spaced energy levels so that they can be characterized, respectively, by the positive frequencies $\omega_c$ and $\omega_h$ as shown in Fig.~\ref{fig:No_Mediator}. The machine cycle consists of the two following strokes.
\begin{enumerate}
	\item One  system $S_c$,  initially at equilibrium at temperature $T_c$, interacts for a time $\tau$ with a system $S_h$, initially at equilibrium at temperature $T_h$.
	\item  After the interaction or collision, $S_c$ and $S_h$ thermalize again.
\end{enumerate}

In realistic implementations of this kind of thermal machines, there could also be  other times required, for example, to re-initialize the machine. We denote by $t_w$ the waiting time between the end of a collision and the start of the following one. Then, each cycle lasts $\tau +t_w$. Depending on the physical realization, the time $t_w$ can vary over a wide range, from being negligible compared to $\tau$ to being much larger than that. For example, Fig.~\ref{fig:No_Mediator} describes a situation similar to that of Refs.~\cite{Allahverdyan2010,Campisi2015,Molitor2020Stroboscopic}, in the case each collection is made of only one system and the interaction time between the machine systems is assumed to be negligible compared to the waiting time, which at least comprises the thermalization time. In the opposite case, we assume that both collections are made of many systems and the interaction between couples of systems is activated sequentially. Therefore, if there are enough couples, the systems of the first couple are already practically thermalized when the interaction of the last couple ends~\footnote{Strictly speaking, a system requires an infinite time to thermalize at the temperature of the heat bath it is weakly interacting with~\cite{BookBreuer2007}. However, after a certain finite time, the system reaches a state which is practically indistinguishable from a thermal state.}.  Even in the latter case, the waiting time $t_w$ can not be deemed exactly zero since an unavoidable amount of time may be required for the machine between the end of a collision and the beginning of the following one. Having in mind this scenario, in our analysis we consider $t_w$ as a parameter which can vary. 

We notice that a model sharing some similarity with the one we propose in this section has been very recently investigated~\cite{Molitor2020Stroboscopic}. There, differently from here, the time allocated for the second stroke has a different physical origin being it connected to partial thermalization by means of a collisional bath.

\begin{table*}[t]
	\centering
	\resizebox{\textwidth}{!}{%
		\begin{tabular}{|c|c|c|c|c|}
			\hline
			& Engine  & Refrigerator & Heat pump & Thermal accelerator \\ \hline
			Conditions 
			& $\omega_c < \omega_h < \omega_c \prt{T_h/T_c}$ 
			& $ \omega_h > \omega_c \prt{T_h/T_c}$ 
			& $ \omega_h > \omega_c \prt{T_h/T_c}$ 
			& $ \omega_h < \omega_c $ \\ \hline
			Work and heat
			& $W < 0$, $Q_h > 0$, $Q_c < 0$
			& $W > 0$, $Q_h < 0$, $Q_c > 0$
			& { $W > 0$, $Q_h < 0$, $Q_c > 0$}
			& $W > 0$, $Q_h > 0$, $Q_c < 0$ \\ \hline
			Efficiency or COP 
			& $-W /Q_h =1 - \omega_c/\omega_h$ 
			& $ Q_c / W  =\omega_c/\prt{\omega_h - \omega_c}$ 
			& $-Q_h/W = \omega_h/\prt{\omega_h - \omega_c}$ 
			& $-Q_c/W = \omega_c/\prt{\omega_c - \omega_h}$ \\ \hline
		\end{tabular}%
	}
	\caption{	Working regimes of the cycle. The conditions of the first line are valid when $N_c=N_h$ and $\min(\ev{\nc}_\thrm,\ev{\nh}_\thrm) \leq \ev{\nr}_{\tau} \leq \max(\ev{\nc}_\thrm,\ev{\nh}_\thrm)$. Regarding the cycle with mediator (see Sec.~\ref{sec:CycleMediator}), we also require that $N_c={N_m}$ and that $\min(\ev{\ns}_{u_r-1},\ev{\nr}_\thrm) \leq \ev{\ns}_{u_r} \leq \max(\ev{\ns}_{u_r-1},\ev{\nr}_\thrm) \ \forall \ u_r > 0$, where $N_m$ is the dimension of the mediator,  $\ns$ its number operator, and $u_r$ the number of its collisions with systems $S_r$  (see Appendix~\ref{APPsec:EfficienciesMediator} for more details). Notice that the efficiencies $\eta$ and the COP do not depend on the dimensions of the systems and the conditions on the time evolution.
	}
	\label{table:MachineFrequencies}
\end{table*}

The free Hamiltonians of the systems of each collection  are given by
\begin{equation}
\label{eq:SystemsHamiltonian}
H_r = \hbar \omega_r \nr, \quad \nr =  \ard\ar,
\end{equation}
where $r=c,h$, depending on which kind of system we are describing. The operators $\ar$ and $\ard$ satisfy
\begin{align}
\label{eq:LadderOperators}
&\ar \ket{n} = \sqrt{n} \ket{n-1},
\quad
&\ard& \ket{n} = \sqrt{n+1} \ket{n+1},\nonumber \\
&\ar\ket{0} = 0,
\quad
&\ard& \ket{N_r -1} = 0.
\end{align}
The use of the operators $\ar$ and $\ard$ allows us to have a unified formalism whatever the Hilbert space dimensions of the systems are, including qubits and harmonic oscillators as special cases.
The commutator of $\ar$ and $\ard$ gives
\begin{equation}
\label{APPeq:CommutatorLadderOperators}
\comm{\ar}{\ard} = \mathbb{I} - d_r \dyad{d_r -1},
\end{equation}
where $\mathbb{I}$ is the identity operator in the Hilbert space of dimension $N_r$, $d_r=N_r$ for any finite-dimensional case (e.g, the qubit case is obtained by choosing $d_r=N_r=2$), while the correct result for the harmonic oscillator case ($\comm{\ar}{\ard} =\mathbb{I}$) is formally obtained by putting $d_r=0$.

Regarding the interaction Hamiltonian for the collisions $S_c$-$S_h$, we use
\begin{equation}
\label{eq:InteractionHamiltonian}
H_{I} = \hbar g \prt{a_c^\dagger a_h + a_c a_h^\dagger},
\end{equation}
where, without loss of generality, the coupling constant $g$ is assumed to be real and positive.
This Hamiltonian, which can be described as \enquote{exchange} or \enquote{tight-binding} in solid state and condensed matter physics and as \enquote{beam-splitter} in quantum optics, is widely used in the literature~\cite{BookHaroche2006,BookBreuer2007}, for example, in the case of qubits~\cite{HewgillPRA2018,ManatulyPRE2019,CakmakPRA2019,heineken2020quantum} and harmonic oscillators~\cite{Reid_2017,DeChiara2018}.
Notice that the Hamiltonian of Eq.~\eqref{eq:InteractionHamiltonian} preserves the total excitation number in the system $S_c + S_h$, i.e.,
\begin{equation}
\label{eq:AssumptionCommutation}
\comm{\nc + \nh}{H_I} = 0.
\end{equation}

During the interaction, systems $S_c$ and $S_h$ are considered isolated from the environments so that the external work $W$, required to switch on and off the interaction, is equal to the difference in energy before and after the collision:
\begin{equation}
\label{eq:WorkNM}
W = \hbar \prt{\omega_h - \omega _c}\prt{\ev{\nh}_{\tau} - \ev{\nh}_{\thrm}},
\end{equation}
where $\ev{\nr}_{\tau}$ and $\ev{\nr}_{\thrm}$ are the average number of excitations of the system  $S_r$, after the interaction and in the thermal state at temperature $T_r$ for the free Hamiltonian of system $S_r$, respectively. In particular, we have $\ev{\nr}_{\thrm}={\rm Tr} \{\rho^\thrm_r \nr\}$,  where we have defined the equilibrium thermal state $\rho^\thrm_r =Z_r^{-1}\exp[-H_r/(k_B T_r)]$, being $Z_r={\rm Tr}\{\exp[-H_r/(k_B T_r)]\}$ the partition function and $k_B$ the Boltzmann constant. 
Eq.~\eqref{eq:WorkNM}  clearly shows that $W$ can be different from zero only when $S_c$ and $S_h$ are not resonant.  We also notice that the number-conserving property of the interaction Hamiltonian implies $\ev{\nh}_{\tau} - \ev{\nh}_\thrm=\ev{\nc}_\thrm-\ev{\nc}_{\tau} $.  We use the convention that the external work $W$ is negative in the case of work extraction. 

After the interaction, the system $S_r$ (with $r=c,h$) is put in contact with its thermal bath at temperature $T_r$ and thermalizes again. In this case, the system exchanges exclusively heat with its bath equal to
\begin{equation}
\label{eq:HeatDefinition}
Q_r = \hbar \omega_r \prt{\ev{\nr}_{\thrm}-\ev{\nr}_{\tau}}.
\end{equation}

An engine is realized when $W < 0$, a refrigerator or heat pump when $W > 0$ and $Q_c > 0$ while a thermal accelerator \footnote{In literature, a thermal accelerator is considered to be a machine which facilitates the natural heat flow from the hot bath to the cold bath.} is obtained for $W > 0$ and $Q_c < 0$, as summarized in Table~\ref{table:MachineFrequencies}. The consequence of choosing an interaction Hamiltonian preserving the total excitation number in the system $S_c + S_h$ [cf. Eq.~\eqref{eq:AssumptionCommutation}] is that the efficiencies and the coefficients of performance (COPs) of these thermal machines only depend on the frequencies of the systems $S_c$ and $S_h$, as reported in Table~\ref{table:MachineFrequencies}. Moreover, under the assumption that $N_c = N_h$ and that  $\min(\ev{\nc}_{\thrm},\ev{\nh}_{\thrm}) \leq \ev{\nr}_{\tau} \leq \max(\ev{\nc}_{\thrm},\ev{\nh}_{\thrm})$\footnote{In the case of qubits and harmonic oscillators, we have verified that this condition and the analogous ones in the case with the mediator [see caption of Table~\eqref{table:MachineFrequencies}] are satisfied for Eq.~\eqref{eq:InteractionHamiltonian}.}, the working regimes depend only on frequencies and temperatures. In fact, it  suffices to notice that, when $N_c = N_h$, 
\begin{equation}
\frac{\omega_c}{T_c} > \frac{\omega_h}{T_h} \iff \ev{\nc}_{\thrm} < \ev{\nh}_{\thrm}.
\end{equation}
Notably, the efficiencies and the COPs here are the same of the \enquote{equivalent} adiabatic Otto cycle~\cite{Cakmak2019,Abah2020}, described in  Sec.~\ref{sec:ComparisonOttoCycle}. We remark that Eqs.~\eqref{eq:WorkNM} and \eqref{eq:HeatDefinition} and the results reported in Table~\ref{table:MachineFrequencies} hold good for any interaction Hamiltonian satisfying Eq.~\eqref{eq:AssumptionCommutation}.

\subsection{\label{subsec:PowerMaximizationNM}Maximization of power}

Here, we solve the problem of maximizing the power of the machine. We remark that all the maximizations in this paper are performed at \ temperatures  and coupling fixed. As a first step, we need to find the average number of excitations in the systems after the collision.
This computation, performed in the Heisenberg picture, is detailed in Appendix~\ref{APPsec:CalculationCollision} and gives for a qubit-qubit collision or an oscillator-oscillator one
\begin{equation} \label{eq:OneCollision0}
\ev{\nh}_{\tau} = \ev{\nc}_{\thrm} + \prt{\ev{\nh}_{\thrm} - \ev{\nc}_{\thrm}}A,
\end{equation}
with
\begin{equation}
\label{eq:OneCollision}
A = \frac{2 \delta^2 + g^2 \prtq{1+\cos\prt{2 k \tau}}}{2 k^2} = 1 - \frac{g^2}{k^2} \sin^2\prt{k \tau}, 
\end{equation}
where $\delta = \prt{\omega_h - \omega_c}/2$ and $k = \sqrt{\delta^2 + g^2}$.
Since the above formulas are exact only in the cases of qubit-qubit and oscillator-oscillator collisions, every detailed analysis in this paper will be done for these cases.
However, in all the other cases, we expect these formulas to be a good approximation when some appropriate conditions are fulfilled (see Appendix~\ref{APPsec:CalculationCollision} for more details).


By using Eqs.~\eqref{eq:OneCollision0} and \eqref{eq:OneCollision} in Eqs.~\eqref{eq:WorkNM} and \eqref{eq:HeatDefinition}, it is possible to cast the functional dependence of the power for the various working regimes into a product of a function $f(T_c,T_h,\omega_c,\omega_h$) that depends solely on temperatures and frequencies and a coefficient $V$:
\begin{align}
\label{eq:PowerFormula}
P_E &= \frac{-W}{\tau + t_{w}} = \hbar \prt{\omega_h - \omega_c} \prt{\ev{\nh}_{\thrm} - \ev{\nc}_{\thrm}} V,\nonumber\\
P_R &= \frac{Q_c}{\tau + t_{w}} = \hbar \omega_c \prt{\ev{\nc}_\thrm - \ev{\nh}_{\thrm}} V, \nonumber \\
P_H &= \frac{-Q_h}{\tau + t_{w}} = \hbar \omega_h \prt{\ev{\nc}_\thrm - \ev{\nh}_{\thrm}} V, \nonumber \\
P_A &= \frac{-Q_c}{\tau + t_{w}} = \hbar \omega_c \prt{\ev{\nh}_\thrm - \ev{\nc}_{\thrm}} V, 
\end{align}
where
\begin{equation}\label{eq:Vterm}
V = \frac{1-A}{\tau + t_w}=\frac{g^2}{k^2}\frac{ \sin^2\prt{k \tau}}{\tau + t_w}.
\end{equation}

After a first study which applies to all the functioning regimes [see Sec.~\eqref{subsec:MaximizationFixedFrequencies}], we will deal with the general problem by focusing on the engine regime [see Sec.~\eqref{subsec:Maximizationwithrespecttofrequencies} and Sec.~\eqref{subsec:MaximizationwithrespecttofrequenciesHO}]. In the case of qubits, we have checked that the qualitative results are the same when studying the figure of merit for refrigerators, which is the product of the cooling power $P_R$ and of the COP, and is the quantity one usually tries to maximize when optimizing a refrigerator cycle~\cite{Allahverdyan2010,Tomas2012}.

\subsubsection{\label{subsec:MaximizationFixedFrequencies}Maximization with fixed frequencies}

We start our investigation on power maximization by studying the case in which both the temperatures and the frequencies of the systems are fixed, as commonly done in the literature (see, for example, Refs.~\cite{Allahverdyan2010,Campisi2015,Abah2017,Cakmak2019,Abah2020}). We also consider the coupling $g$ and the waiting time $t_w$ as given parameters, and the optimization is performed only with respect to the collision time $\tau$. In the next subsection, instead, the optimization will be done also with respect to the frequencies of the systems.  Here, since  $\tau$ only appears in the term $V$, we can maximize all the power functions at once by just maximizing $V$. Notice that when the frequencies are fixed, maximizing the cooling power is equivalent to maximizing the figure of merit since the COP only depends on the frequencies. With this first analysis, we show how much impact the waiting time has on the maximization of power and the optimal collision time for which it is realized.

\begin{figure}
	\centering
	\includegraphics[width=0.48\textwidth]{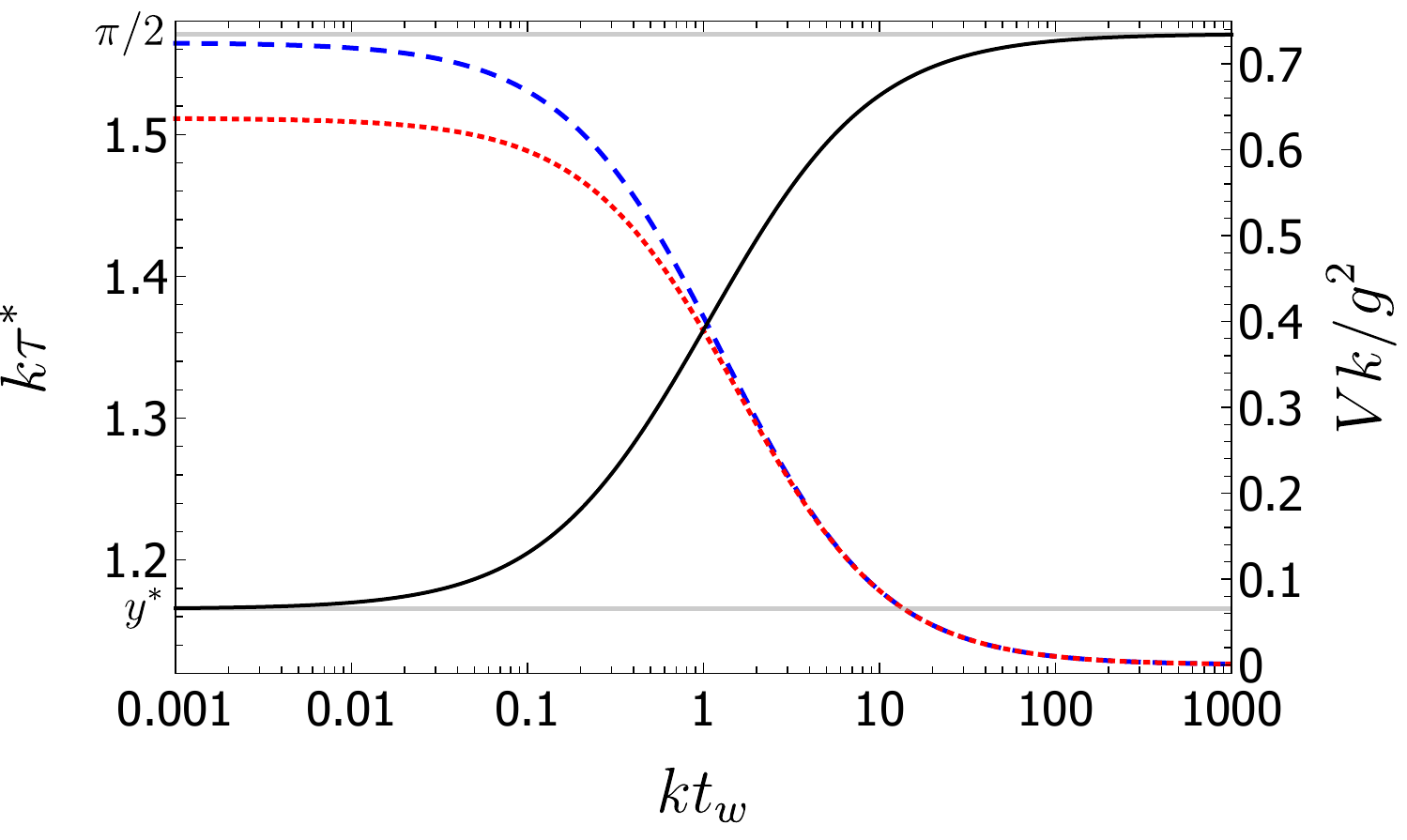}
	\caption{Optimal collision time $\tau^*$ (black continuous curve), $\Vm$, i.e., $V$ calculated at the optimal time $\tau^*$ (blue dashed curve), and $V (k\tau=\pi/2)$, i.e., $V$ calculated at the swap time (red dotted curve), as functions of the waiting time $t_w$ for fixed frequencies and coupling. Each curve presents its greatest variation  when $k t_w$ varies in the interval $[0.1,10]$. Regarding $\tau^*$, the two limit values for $k t_w \rightarrow 0, \infty$ are, respectively, $y^*$ and $\pi/2$.}
	\label{fig:graphOptimalTimeNoMediator}
\end{figure}

The maximization of the term $V$ of Eq.~\eqref{eq:Vterm} with respect to the collision time $\tau$ has to be carried out numerically. In Fig.~\ref{fig:graphOptimalTimeNoMediator}, we report the optimal value of $k \tau$ as a function of the quantity $k t_w$. We denote the optimal collision time as $\tau^*$ and the term $V$ calculated for this collision time as $\Vm$. Moreover, the same plot also shows the value of $V k / g^2$, which is a function of $k \tau$ and $k t_w$, in two cases: for $\tau= \tau^*$ and $k \tau = \pi/2$. The first case is the value of $\tau$ maximizing the power while the second one is the smallest value that maximizes the work per cycle (this happens for $k \tau = \pi/2+ m\pi$, where $m$ is a natural number). When $t_w = 0 $, the optimal collision time is given by the equation $k \tau^* = y^*$, where $y^* \simeq 1.16556$ is the value of $y$ maximizing $\sin^2(y)/y$, the maximum being equal to $\alpha \simeq 0.724611$.
On the other hand, in the limit $t_w \rightarrow \infty$, the maximum of $V$ is simply obtained for $k \tau= \pi /2$ since the contribution of $\tau$ in the denominator of Eq.~\eqref{eq:Vterm} becomes negligible in this limit.
We also denote the time value for which $k \tau = \pi/2$ as the swap time since the states of the two systems $S_c$ and $S_h$ are practically swapped when $g \gg \abs{\delta}$. Figure~\ref{fig:graphOptimalTimeNoMediator} shows that the optimal time drastically changes value when $k t_w$ varies in the interval $[0.1,10]$ and that up to $k t_w \sim 10$ the dimensionless optimal time $k \tau^*$ is	 substantially different from $\pi/2$. However, the advantage in power obtained by using the optimal time instead of the swap one is significant only up to $k t_w \sim 0.1$.

\begin{figure}
	\centering
	\includegraphics[width=0.48\textwidth]{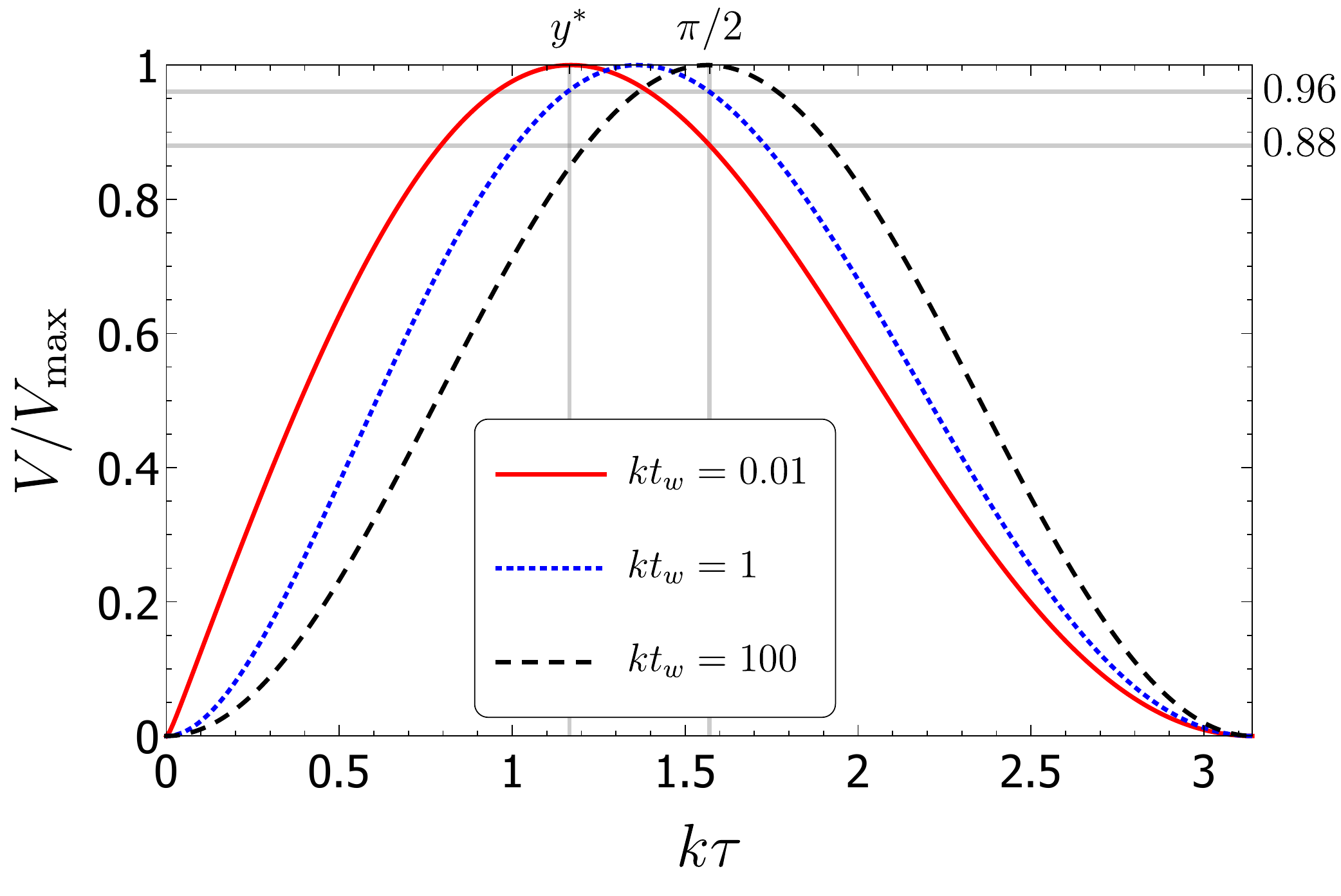}
	\caption{$V/\Vm$ as a function of the collision time $\tau$ for fixed frequencies and coupling. In this case, the ratio is only a function of $k \tau$ and $k t_w$. Small errors in the collision time around $\tau^*$ provoke only a small decrease of $V$.}
	\label{fig:graphTimesNoMediator}
\end{figure}

Remarkably, the power output is quite resilient to errors in the interaction time, as can be seen in Fig.~\ref{fig:graphTimesNoMediator}. There, we show how $V/\Vm$ changes as a function of $k \tau$. Small uncertainties in the collision time around the optimal collision time $\tau^*$ lead to a small decrease in the output power. The plot also shows that the loss in power obtained by using the swap time instead of the optimal one increases when $t_w$ decreases, becoming approximately $12\%$ for $k t_w=0.01$.

The perfect swap between two non-resonant systems (e.g., qubits in Ref.~\cite{Campisi2015}) is often assumed to be a valid operation in a thermodynamic cycle. One could wonder about the power of a Hamiltonian capable of exactly implementing a perfect swap and its performance. We analyze this case in Appendix~\ref{APPsec:PerfectSwap} for the qubits case, finding that the \enquote{swap} Hamiltonian preserves the number of excitations in a collision, thus leading to the same efficiencies of our interaction Hamiltonian of Eq.~\eqref{eq:InteractionHamiltonian}. We compare the performance of the exchange and swap Hamiltonians by imposing that the difference between the highest and lowest eigenvalues is the same in the two Hamiltonians (for simplicity, this is done for $t_w = 0$). We find that the swap Hamiltonian performs better when it is possible to optimize the interaction time, while this is not always true if the swap Hamiltonian acts for the appropriate swap time $t_S$.  For this case, we find for which values of $g$ the exchange Hamiltonian leads to higher power output than the swap one. 

\subsubsection{\label{subsec:Maximizationwithrespecttofrequencies} Maximization with respect to frequencies, qubits case}

Here, we deal with the problem of maximizing the power output not only with respect to the collision time but also with respect to the frequencies of the two collections, assuming that both of them are composed of qubits. Contrary to the previous maximization of the term $V$ at fixed frequencies, maximizing over frequencies impacts the efficiencies of the thermal machine and leads to different results for different working regimes. We choose to make this maximization in the engine case. Similar qualitative results are obtained by studying the figure of merit for refrigerators.

We start our analysis by observing that, for properly optimizing the power, it is not sufficient to maximize with respect to $\omega_c/\omega_h$. Therefore, since we cannot use $\omega_c$ as the unit frequency, we define the frequency $\nu_c = k_B T_c/\hbar$ as a unit measure for frequencies.


\begin{figure}
	\centering
	\includegraphics[width=0.48\textwidth]{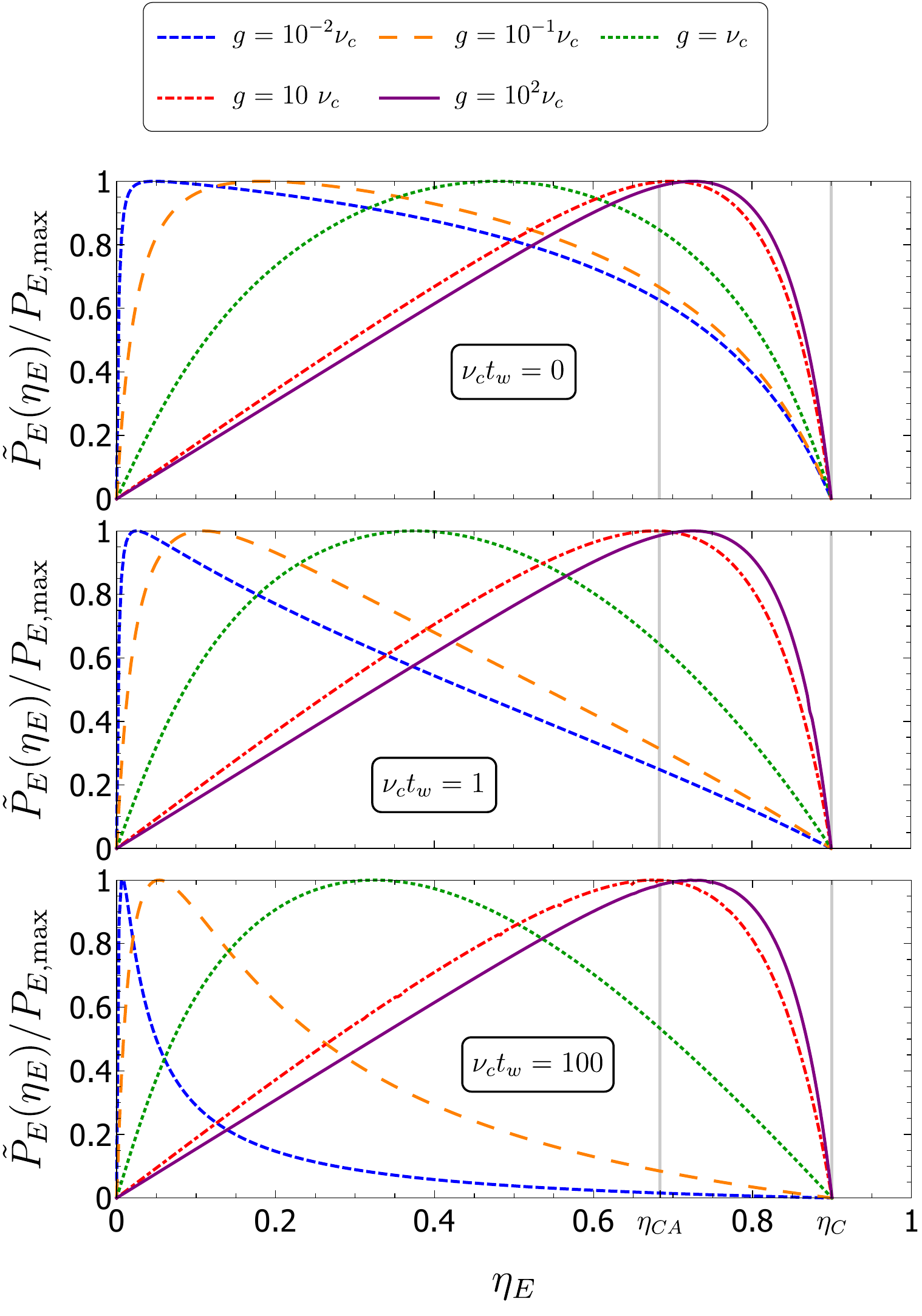}
	\caption{Normalized maximum power $\tilde{P}_E (\eta_E)/P_{E, \mathrm{max}}$ as a function of the efficiency $\eta_E=1-\omega_c/\omega_h$. For a given $\eta_E$, the maximization is done with respect to $\tau$ and $\omega_c$, imposing $\omega_h = \omega_c/(1-\eta_E)$. Each plot shows the same curve for different values of the coupling $g$, while they differ for the value of $t_w$. Other parameters are: $N_c=N_h=2$ and $T_h = 10 T_c$. The two vertical lines indicate, from left to right, the Curzon-Ahlborn efficiency and the Carnot one.	Noticeably, in the high coupling limit, the efficiency at maximum power is higher than the Curzon-Ahlborn one for every waiting time.}
	\label{fig:effVsPowGraphTotalMax}
\end{figure}

In Fig.~\ref{fig:effVsPowGraphTotalMax}, we show the behavior of the normalized maximum power as a function of the efficiency for different values of the waiting time. Noticeably, the power peaks are characterized by an efficiency larger than the Curzon-Ahlborn one provided a high enough coupling is employed. Therefore, our model belongs to the class of systems that may surpass the Curzon-Ahlborn efficiency at maximum power~\cite{VanDenBroeck2005,Esposito2009,Esposito2010,Shiraishi2016,Shiraishi2017,BookBinder2018,Allahverdyan2010,Campisi2015,Erdman2019Maximum}. Moreover, the curves corresponding to the high coupling limit do not change significantly when increasing the waiting time $t_w$ while the other ones have their peak moving to the left.

We would like to comment on what the high coupling limit represents in this setting, where the frequencies can vary. The high coupling limit is obtained when $g \gg \abs{\delta}$ so that $k\simeq g$.
However, $\abs{\delta} = \abs{(\omega_h - \omega_c)/2}$ so that the magnitude of $\delta$ maximizing the power depends on $T_h/T_c$ even if the temperatures do not directly appear in the term $V$.
Our numerical simulations suggest that the optimal frequencies are roughly comprised in the range $[\nu_c, (T_h/T_c) \nu_c]$ (the upper bound can be slightly overcome for small values of $\eta_E$), hence a safe condition for the high coupling limit is $g \gg (T_h/T_c) \nu_c$.

\subsubsection{\label{subsec:MaximizationwithrespecttofrequenciesHO} Maximization with respect to frequencies, harmonic oscillators case}

In the case of harmonic oscillators, the power peak is not obtained for some specific finite non-zero values of $\omega_c$ and $\omega_h$, as shown
in Appendix~\ref{APPsec:PowerHarmonicOscillators}. There, the power is indeed shown to increase monotonically by decreasing the frequencies while keeping the temperatures and the efficiency fixed.

When searching for the maximum power, since it is obtained in the limit $\omega_c, \omega_h \rightarrow 0$, the high coupling condition ($g \gg |\delta|$) is always respected so that $\Vm$ is independent from the frequencies for any finite coupling [see Eq.~\eqref{APPeq:ktermOmegac}]. For the same reason ($\omega_c, \omega_h\rightarrow 0$), the search for the maximum can be done in the mathematically equivalent high-temperature limit ($k_B T_r \gg \hbar \omega_r$).
To find $P_E$ of Eq.~\eqref{eq:PowerFormula} in this case, let us write $\omega_h =\omega_c/(1-\eta_{E})$ and  $T_h = T_c / (1 - \eta_C)$, where $\eta_C$ is the Carnot efficiency.
Then, after some straightforward calculations, we obtain
\begin{equation}
\label{eq:HighTemperaturesHarmonicOscillatorPower}
P_E \simeq \frac{k_B T_c}{1-\eta_C} \frac{\eta_{E} \prt{\eta_C-\eta_{E}}}{1-\eta_{E}} V.
\end{equation}
When maximizing $P_E$ in this limit, we recall that $\Vm$ is independent of the frequencies and, consequently, of the efficiency $\eta_{E}$.
Then, one can just maximize $[\eta_{E} (\eta_C-\eta_{E})]/(1-\eta_{E})$ in Eq.~\eqref{eq:HighTemperaturesHarmonicOscillatorPower} finding that the efficiency maximizing the power is the Curzon-Ahlborn one.

The distinct behavior between qubits and harmonic oscillators stems from the very different probability distribution of the populations in the thermal states, which, being both systems characterized by a single frequency, is entirely due to the difference in the number of levels. It follows that although the Heisenberg dynamics of the observables we are interested in during the collision is the same, a maximization over frequencies leads to different results. In particular, the monotonic increase of power by decreasing the frequencies  in the case of oscillators appears to be connected to their capacity to absorb an unlimited number of excitations in the limit of vanishing frequencies, in contrast with finite dimensional systems.

\section{\label{sec:CycleMediator}The model with mediator}

As we have seen in the previous section, when the number of couples $S_c$-$S_h$ at disposal is large enough, the contribution of the thermalization to the effective waiting time $t_w$ can be eliminated, greatly reducing $t_w$, whose final value will depend, for example, on the time required by the machine between a collision and the next one. However, for a small number of couples, the relaxation time cannot be neglected. In this situation, one could try to improve the setup by adding a mediator system $S_m$ which alternately interacts with one of the two systems $S_c$ and $S_h$ in order to double the number of collisions, thus effectively extending the time allocated for the thermalization. This can practically eliminate the contribution of the thermalization time to the effective waiting time $t_{w, m}$ for a relatively large number of couples even when this number is not sufficient for the case without the mediator. Moreover, the addition of a mediator could also modify the part of the waiting time due to machine requirements. For example, if some amount of time is required between subsequent collisions, with the presence of the mediator this time could be suppressed because the interaction $S_m$-$S_h$ can be turned on as soon as the $S_m$-$S_c$ is turned off.

\begin{figure}
	\centering
	\includegraphics[width=0.4\textwidth]{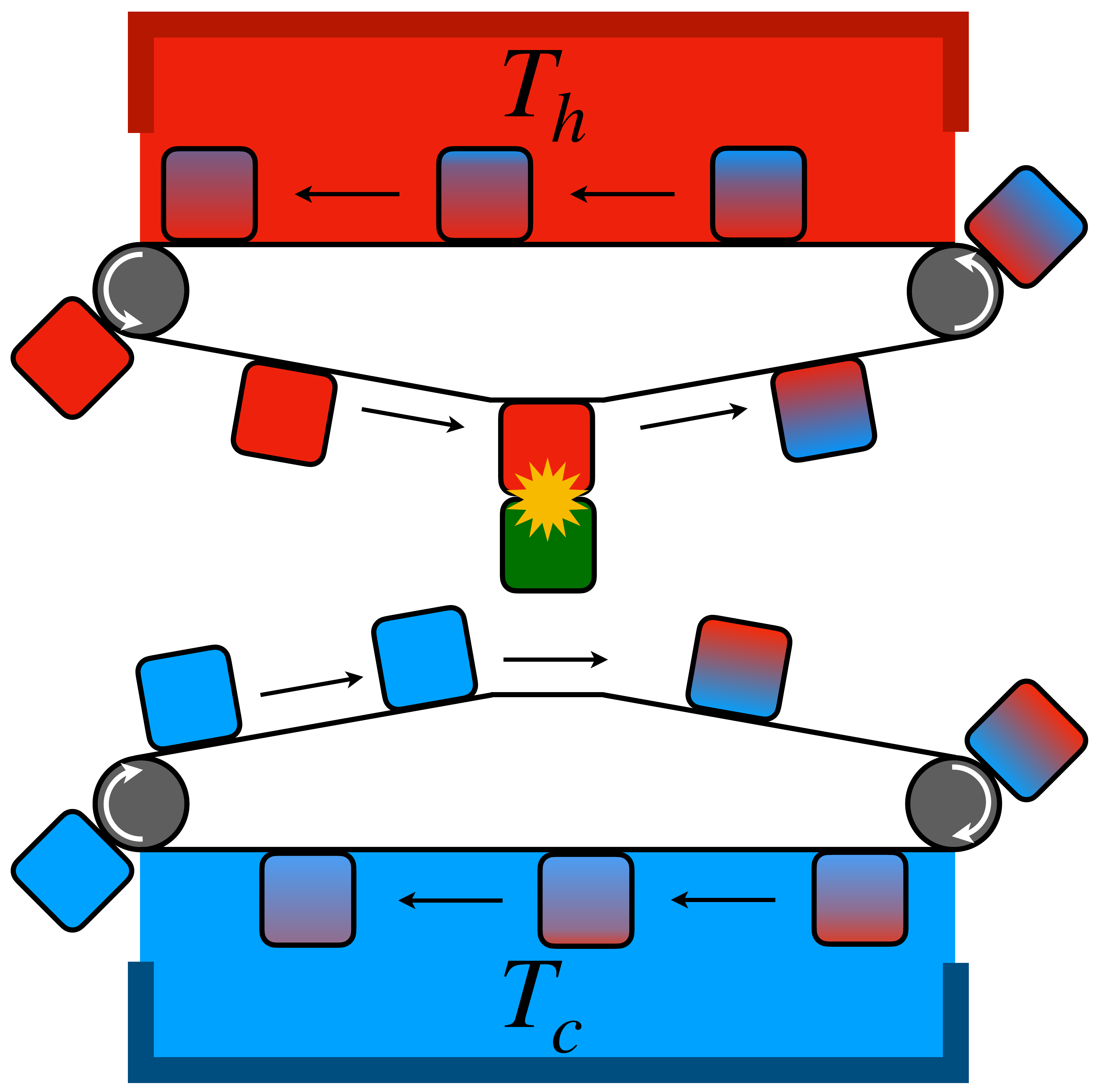}
	\caption{A scheme of the cycle with the mediator. Here, for each cycle,  the mediator system interacts alternately with a system of collection $C_c$ and then with a system of $C_h$ (in the general case considered in the text, one can have $u_r$ collisions).}
	\label{fig:CycleSchemeMediator}
\end{figure}

The new setup can be described as follows: we consider a machine composed of the same components as in the previous section, but with the addition of a central \enquote{mediator} system $S_{m}$ characterized by the frequency $\omega_{m}$. Moreover, we consider the possibility of grouping the systems of each collection in order to make the mediator interact, alternately, with sequences of systems of the same collection. Then, in the most general case, the cycle consists of  the two following strokes.
\begin{enumerate}
	\item  The mediator interacts consecutively with $\uc$ systems of the collection $C_c$, each collision lasting for a time $\tau_c$ through an interaction Hamiltonian like that of Eq.~\eqref{eq:InteractionHamiltonian} but with coupling $g_c$ and operators  $\am$ and $\amd$ [satisfying the same conditions of Eq.~\eqref{eq:LadderOperators}] instead of, respectively, $g$,   $\ah$, and $\ahd$.
	\item  The mediator interacts consecutively with $\uh$ systems of the collection $C_h$, each collision lasting for a time $\tau_h$ through an interaction Hamiltonian like that of Eq.~\eqref{eq:InteractionHamiltonian} but with coupling $g_h$ and operators  $\am$ and $\amd$  instead of, respectively, $g$,   $\ac$, and $\acd$.
\end{enumerate}
This model and its functioning are illustrated in Fig.~\ref{fig:CycleSchemeMediator} in the particular case $\uc=\uh=1$, which we will argue to be the best performing case.
We assume that, before the interaction with $S_{m}$, the state of the systems $S_r$ (where $r=c,h$) is always described by the corresponding thermal state and that the interacting systems are initially not correlated.

Ideally, as said above, the advantage of this scheme compared to the previous one is that the effective waiting time can be reduced even more. In particular, by comparing the two models in the case $\uc=\uh=1$, we can expect that $t_{w, m} \simeq 0$ when $t_w \lesssim \tau_m^*$, where $\tau_m^*$ is the optimal collision time of the model with the mediator (computed in the case $t_{w, m}=0$), since  $t_{w,m} \simeq \max\{0, t_w - \tau_m^*\}$. For simplicity, within this model, we consider the case when one can effectively neglect the waiting time, i.e., when $t_{w, m}=0$.

Besides a possible performance advantage, the addition of a mediator could make the experimental implementation of the model easier. In fact, with the mediator, one needs only to turn on and off the interactions between different systems and a single one, which is always the same, instead of doing it between systems taken from the two collections. For example, moving systems around can present some difficulties in some specific applications. By using the mediator, it is sufficient to move it from one collection to the other instead of always trying to move a different system from each collection.

Assuming excitation-preserving interactions, we obtain the same results for the efficiencies of the previous model (see Table~\ref{table:MachineFrequencies} and Appendix~\ref{APPsec:EfficienciesMediator} for a detailed discussion). If all the systems are of the same type and the conditions on the time evolution due to the collisions reported in Table~\ref{table:MachineFrequencies} are fulfilled, the same conditions on the frequencies for the working regimes are obtained again (see Appendix~\ref{APPsec:EfficienciesMediator}).

\subsection{Maximization of power}

\subsubsection{Maximization with fixed frequencies}

Analogously to the analysis presented in Sec.~\ref{sec:ModelNoMediator}, we keep fixed the two bath temperatures and the coupling constants, and we start by maximizing the power output with respect to the collision time with fixed frequencies.

In Appendix~\ref{APPsec:PowerMaximization} we report the detailed calculations showing that the values of the power for the setup with the mediator are the same of Eq.~\eqref{eq:PowerFormula} with the substitution $V \rightarrow V_m$ where
\begin{equation}
\label{eq:VtermMediator}
V_{m} = \frac{\prtq{1-(A_c)^{u_c}}\prtq{1-(A_h)^{u_h}}}{\prt{u_c \tau_c + u_h \tau_h} \prtq{1 - (A_c)^{u_c} (A_h)^{u_h}}},
\end{equation}
and the quantities $A_r$, with $r=h,c$, are the equivalent of the quantity $A$ of Eq.~\eqref{eq:OneCollision} with the substitutions $g \rightarrow g_r$, $\delta \rightarrow \delta_r = (\omega_r - \omega_m)/2$,  $k \rightarrow k_r = \sqrt{g_r^2 + \delta_r^2}$, and $\tau\rightarrow \tau_r$. Each of these quantities concerns the collisions $S_r$-$S_{m}$ and is analogous to those of the model without the mediator. However, despite this similarity, Eq.~\eqref{eq:VtermMediator} is too difficult to maximize exactly or to obtain a simple general numerical solution as for Eq.~\eqref{eq:Vterm}. To simplify the problem we assume that  $u_c = u_h{ \equiv u_m}$ and $g_c = g_h \equiv g_m$. In this case, numerical simulations show that the maximum of $V_m$ is obtained for $\tau_c = \tau_h \equiv \tau_m$ and $\omega_m = \bar{\omega} \equiv (\omega_c + \omega_h)/2$, leading to $-\delta_c = \delta_h = \mdelta \equiv (\omega_h - \omega_c)/4$. With these assumptions, $A_c = A_h \equiv A_m$ and we can cast the term $V_m$ in the simplified form
\begin{equation}
\label{eq:VtermMediatorMoreCollisions}
V_m = \frac{1 - (A_m)^{u_{m}}}{2 u_{m} \tau_m \prtq{1+(A_m)^{u_{m}}}}.
\end{equation}

Numerical simulations show that the maximum power is achieved when system $S_{m}$ makes just one collision with each collection corresponding to $u_{m}=1$ (see Appendix~\ref{APPsec:PowerMaximization} for additional considerations). In this case, $V_m$ further simplifies to
\begin{equation}
\label{eq:VtermEqualCoupling}
V_m = \frac{g_m^2 \sin^2\prt{k_m \tau_m }}{2 \tau_m \prtq{2 k_m^2 - g_m^2 \sin^2\prt{k_m \tau_m}}}.
\end{equation} 
We stress that in the following numerical study we always impose $g_c = g_h = g_m$ and $\tau_c = \tau_h = \tau_m$.

\begin{figure}
	\centering
	\includegraphics[width=0.48\textwidth]{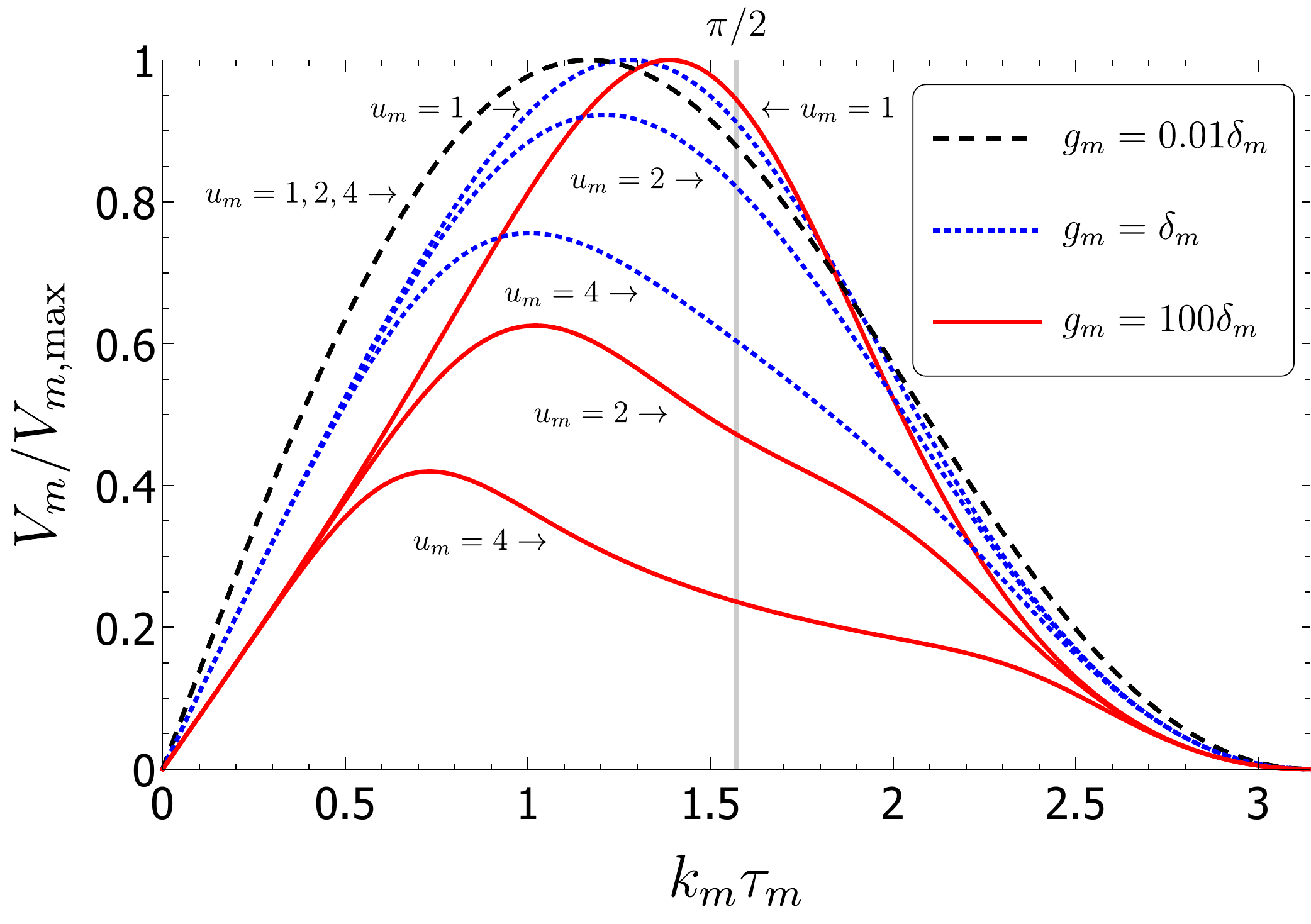}
	\caption{Normalized term $V_m/V_{m, \mathrm{max}}$, $V_{m, \mathrm{max}}$ being the maximum of $V_m$ in the case ${u_m}=1$ (for the same value of $g_m$), as a function of time for fixed frequencies and temperatures. The various curves refer to different values of the coupling $g_m$ with respect to the detuning term $\delta_m = (\omega_h - \omega_c)/4$. Moreover, for each choice of $g_m$, we plot three times the term $V_m$ with $u_m$ assuming the values $1$, $2$, and $4$ (curves with the same $g_m$ are plotted with the same style). Increasing $u_m$ always lowers the power of the machines. When $g_{m} = 0.01 \delta_m$ the three curves practically coincide. We remark that the positions of the peaks move when the ratio $g_m/\delta_m$ varies.}
	\label{fig:graphTimesWithMediator}
\end{figure}

Figure~\ref{fig:graphTimesWithMediator}  shows that, as in the no-mediator case, the protocol is resilient to small errors in the collision times. Differently from the no-mediator case, the optimal value of $k_m \tau_m$ cannot be made independent on $g_m$ and $\delta_m$ because Eqs.~\eqref{eq:VtermMediatorMoreCollisions} and \eqref{eq:VtermEqualCoupling} cannot be written as a single-variable function of $k_m \tau_m$. In fact, Fig.~\ref{fig:graphTimesWithMediator} shows how the peak of $V_m$ moves to the right with increasing $g_m$. Figure~\ref{fig:graphTimesWithMediator} also shows that multiple collisions decrease the performance and that when $g_{m} \ll \abs{\delta_m}$ the loss in power for doing multiple collisions is negligible.

\begin{figure}
	\centering
	\includegraphics[width=0.48\textwidth]{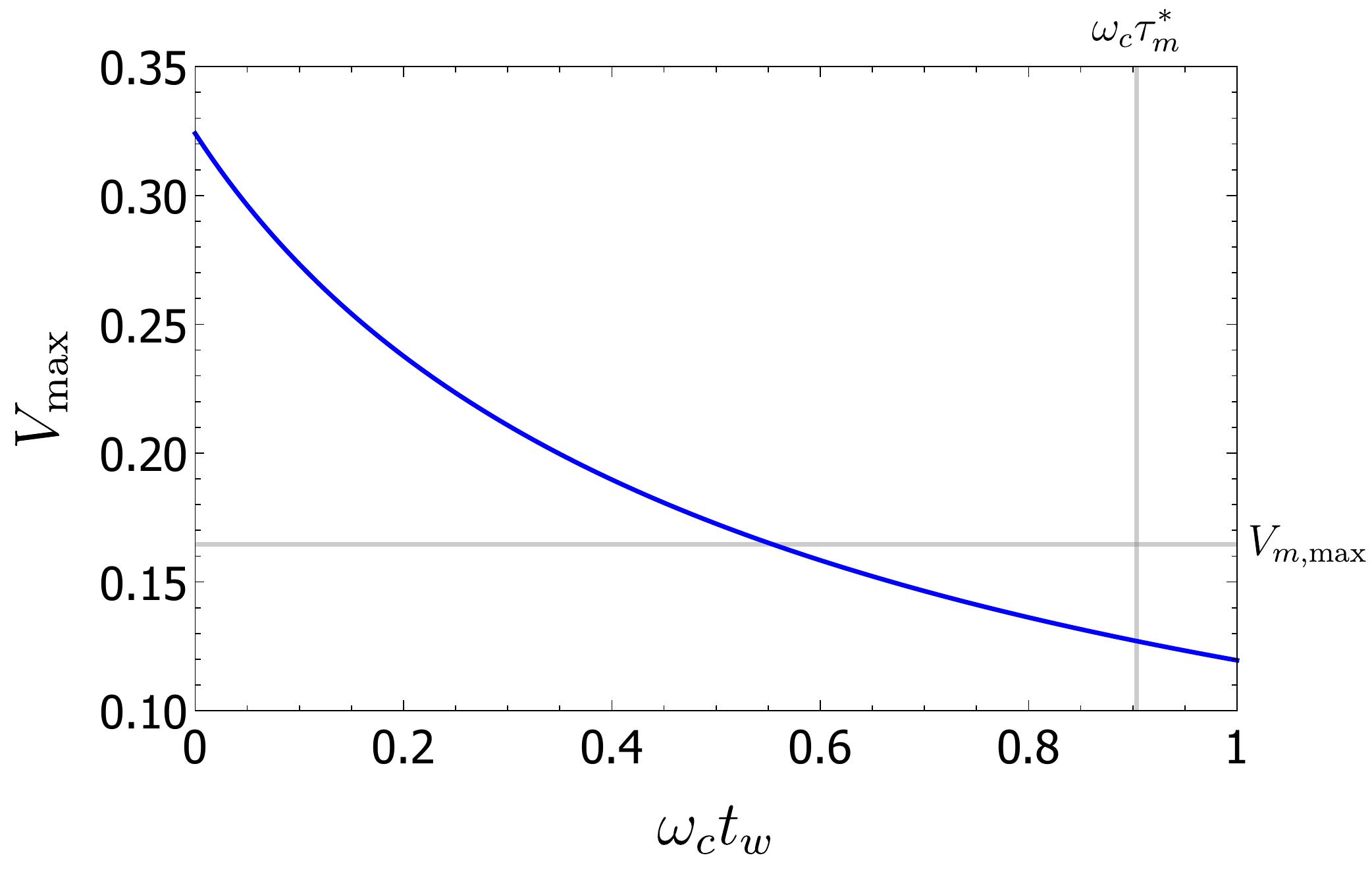}
	\caption{$\Vm$ of the model without mediator against the waiting time $t_w$ with $\omega_h = 5 \omega_c$ and $g=g_m=\omega_c$. The vertical line indicates the optimal interaction time in the model with the mediator, while the horizontal line indicates the term $\VmedMax$. The zone where $\Vm$ is lower than $\VmedMax$ and $t_w\lesssim \tau_m^*$ indicates a power advantage of the model with the mediator compared to the model without it (for $t_w > \tau_m^*$ the assumption $t_{w, m}=0$ is not justified, in general).}
	\label{fig:MediatorAdvantage}
\end{figure}

Hereafter,  we only deal with the case $u_m=1$, which seems to give the best performance. In the two limiting cases of small and large coupling, it is easy to see from Eq.~\eqref{eq:VtermEqualCoupling} that the no-mediator cycle (with $g=g_m$) with $t_w=0$ greatly outperforms the version with the mediator (we recall that $t_{w, m}=0$) in terms of maximum power [cf. Eq.~\eqref{eq:Vterm}]. Moreover, numerical simulations confirm that this remains valid for any value of $g_{m}$. However, the cycle with the mediator can perform better  when $t_w > 0$ (but  $t_{w, m}=0$), as shown in Fig.~\ref{fig:MediatorAdvantage}. There, the term $\Vm$ is plotted against the waiting time $t_w$. We observe a parameter zone where $t_w \lesssim \tau_m^*$ and $\VmedMax > \Vm$ in which the system with mediator performs better than the system without it. We recall that for $t_w > \tau_m^*$ the assumption $t_{w, m}=0$ is not justified, in general.

\subsubsection{Maximization with respect to frequencies, qubits case}

\begin{figure}
	\centering
	\includegraphics[width=0.48\textwidth]{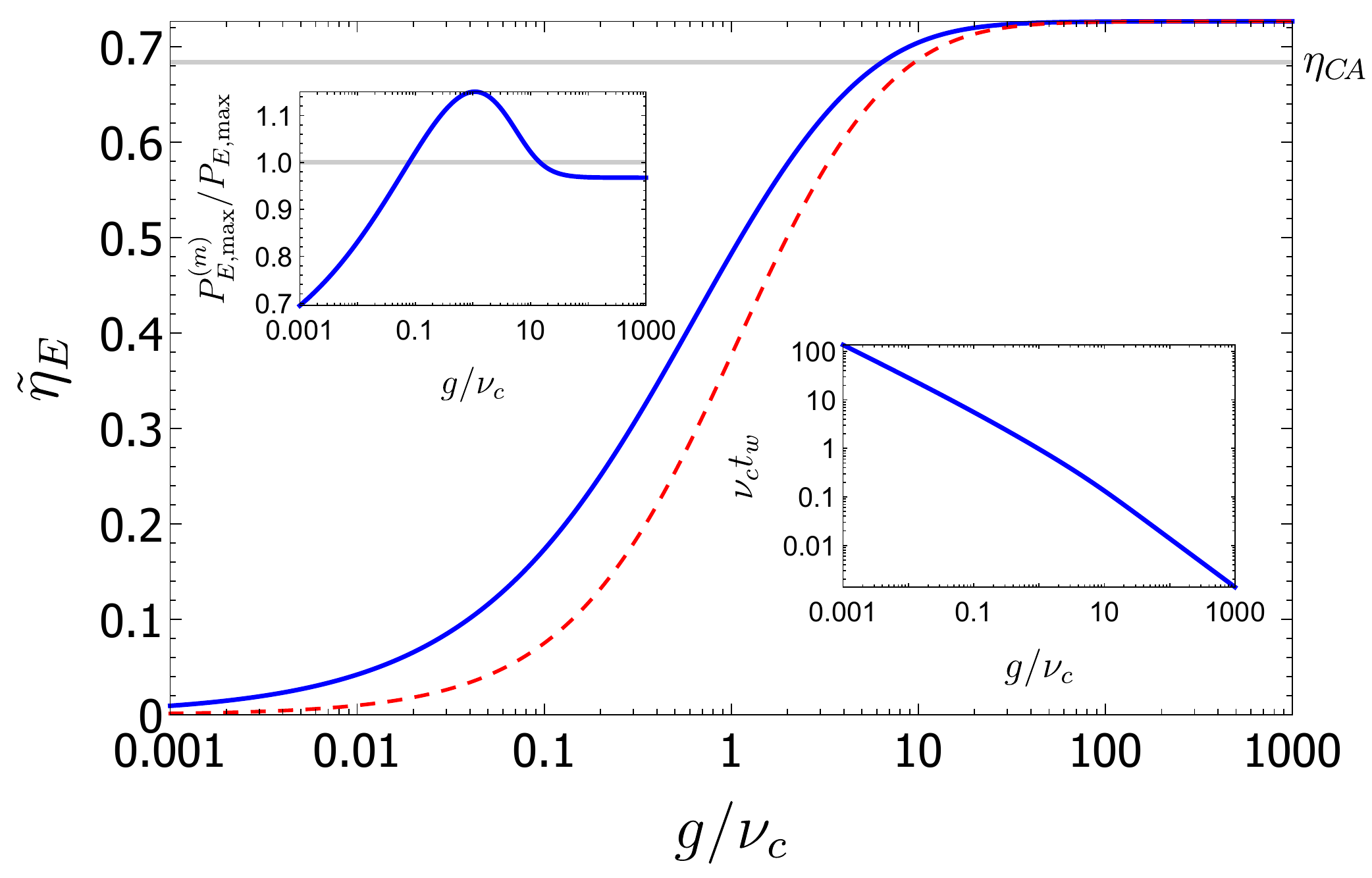}
	\caption{The continuous blue line refers to the efficiency at maximum power, ${\tilde{\eta}_E}$, as a function of the coupling $g$ ($g=g_m$ in this plot) for the cycle with the mediator. The maximization is over frequencies and collision time. The frequency $\nu_c = k_B T_c/\hbar$ is the measure unit and $T_h = 10 T_c$.    The horizontal line indicates the Curzon-Ahlborn efficiency for the given temperatures. The red dashed line refers to the efficiency at maximum power in the same case of the continuous blue line but for the cycle without the mediator and $t_w = \tau_m^*$. The top-left inset shows the ratio of the maximum power in the cycle with the mediator over the maximum power in the cycle without the mediator (here too with $t_w = \tau_m^*$). The bottom-right inset shows $t_w$ or, equivalently for this plot, $\tau_m^{*}$ as a function of the coupling $g$. The graph shows that there can be performance advantages in using the mediator when $t_{w} \sim \tau_m^*$.}
	\label{fig:MediatorAdvantageMaximumPower}
\end{figure}

In the previous section, we have seen with a specific example that there are configurations in which the cycle with the mediator can offer performance advantages compared to the cycle without it. However, we have observed this behavior by choosing specific frequencies for the collections. In this subsection, we show that the advantage remains if we maximize both cycles not only with respect to the collision time but also over the frequencies. To do so, we focus on the engine working regime in the case in which all the systems are qubits.

Figure~\ref{fig:MediatorAdvantageMaximumPower} shows the variation of the efficiency in the optimal configuration of frequencies and  collision times  as a function of the coupling  $g=g_m$, for the fixed temperature ratio $T_h/T_c = 10$. In this plot, the waiting time of the cycle without the mediator is assumed equal, for each value of $g$, to the optimal time of the cycle with the mediator, i.e., $t_w=\tau_m^*$. Under this assumption, the efficiency of the system with the mediator is always  larger than or equal to that of the cycle without it. Regarding the power output, the top-left inset shows the ratio of the two power maximum outputs ($P^{(m)}_{E,\textrm{max}}$ is the maximum power in the model with the mediator), $P^{(m)}_{E,\textrm{max}}/P_{E,\textrm{max}}$. We can see that around $g= \nu_c$ the cycle with the mediator also outputs more power than the cycle without it, thus providing a complete performance advantage. The bottom-right inset shows the optimal time of the collision in the cycle with the mediator (and thus the waiting time used for the cycle without the mediator) as a function of the coupling.

\begin{figure}
	\centering
	\includegraphics[width=0.48\textwidth]{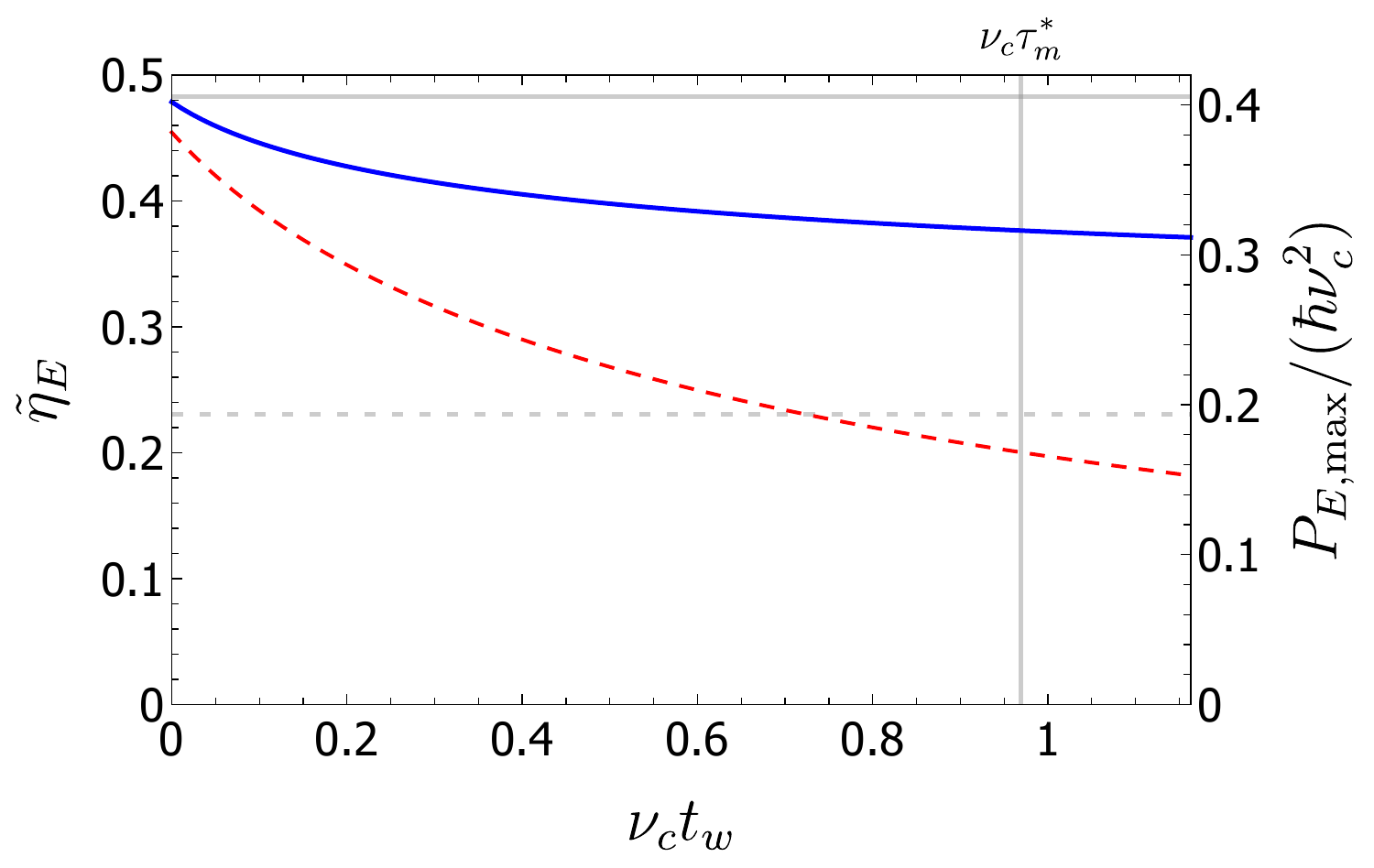}
	\caption{Efficiency at maximum power ${\tilde{\eta}_E}$ (continuous blue line) and maximum power $P_{{E,}\textrm{max}}$ (dashed red line) of the model without the mediator as a function of the waiting time $t_w$ with $T_h= 10 T_c$ and $g=g_m=\nu_c$ (we recall that $t_{w, m}=0$). The vertical line indicates the optimal collision time in the model with the mediator, while the horizontal lines indicate, respectively, the efficiency at maximum power (continuous line) and the maximum power (dashed line) in the same model. The plot shows that the model with the mediator can have performance advantages.}
	\label{fig:MediatorAdvantageFixedTemperaturesAndCoupling}
\end{figure}

In Fig.~\ref{fig:MediatorAdvantageFixedTemperaturesAndCoupling} we analyze more in detail the case with $g={g_m=}\nu_c$ by letting $t_w$ vary. Regarding the efficiency at maximum power, the cycle with the mediator provides an appreciably higher value provided that the waiting time in the cycle without the mediator is not negligible. Regarding the power, instead, the waiting time $t_w$ has to be a significant fraction of  $\tau_m^* $ to provide an advantage. As before, it must be taken into consideration that for $t_w > \tau_m^*$ the assumption $t_{w, m}=0$ is not justified, in general.

\section{\label{sec:ComparisonOttoCycle}Comparison with the quantum Otto cycle}

In this section, we compare the power of the cycle we have studied in Sec.~\ref{sec:ModelNoMediator} with that of another cycle intensely studied in the literature, the so-called quantum Otto cycle~\cite{Geva1992,Scully2002,Quan2007,Kosloff_2017}. This cycle is composed of four strokes: an expansion, a compression one, and two isothermal strokes. Usually, the system Hamiltonian is changed so that the separation of its eigenenergies increases  during the compression, while it decreases during the expansion.

To compare it with our model without the mediator, we denote $\hbar\omega_c$ the energy separation of the eigenstates of the system when it is in contact with the cold thermal bath and $\hbar\omega_h$ when in contact with the hot thermal bath. If the expansion and compression strokes can be performed adiabatically (requiring then an infinite time), the efficiency of the Otto cycle is given by the same formulas of Table~\ref{table:MachineFrequencies} for engines and refrigerators~\cite{Quan2007,Cakmak2019,Abah2020}. However,  in real applications, the expansion and compression strokes last a finite driving time so that the cycle cannot be performed perfectly adiabatically: the states of the system at the end of the expansion and compression strokes are not the same as in the adiabatic case. Therefore, the power output and the efficiency decrease. To counter this limitation, various techniques called shortcuts to adiabaticity (STA) have been applied during finite-time expansion and compression stages to maintain constant the populations of the energy eigenstates~\cite{Abah2017,Cakmak2019,Abah2020}. However, the STA require an additional cost which increases as the driving time decreases~\cite{Abah2017,Cakmak2019,Abah2020, ZhengPRA2016,CampbellPRL2017,Torrontegui2017,AbahNJP2019}. In the two Otto cycles we consider below, the cost for the STA implementation is computed differently in the two cases, according to the original works, which can be found, respectively, in Refs.~\cite{Cakmak2019} and \cite{Abah2020}.

The same efficiency of the ideal Otto cycle has been found also in other cycles  comprising working substances made of non-resonant components, see for example \cite{Campisi2015,DeChiara2018,DeChiara2020}.
Before comparing the Otto cycle and our cycle without the mediator we comment on what could physically motivate the fact that the ideal Otto cycle and our cycle have the same efficiency. As shown in Sec.~\ref{sec:ModelNoMediator}, the efficiency of our cycle only depends on the frequencies $\omega_c$ and $\omega_h$. If we consider the limit in which the interaction $g$ goes to infinity and $g\tau = \pi/2$, the collision gives place to a swap between the particles. Therefore, the state of the system belonging to collection $C_c$ goes to the system of collection $C_h$ and viceversa. This is equivalent to performing at once the expansion and compression strokes adiabatically because, in this case, the states remain unaltered while the Hamiltonian changes. In our case, instead, the two Hamiltonians are not altered but the corresponding systems swap the states. After the collision, in our cycle, both systems thermalize, thus performing at once the equivalent of the two isothermal strokes of the Otto cycle. To conclude, since there is a region of parameters in which the ideal Otto cycle and our cycle give rise to the same dynamics for the states, they give rise to the same efficiency. Then, since this efficiency, in our cycle, is dependent only on the frequencies, the equality holds also when the two models are not so directly correspondent. 
Overall, the connection between the two models can be then linked to the conversion or exchange of excitations through  the Hamiltonian of Eq.~\eqref{eq:InteractionHamiltonian} between non-resonant systems at frequencies $\omega_c$ and $\omega_h$. 

\begin{figure}
	\centering
	\includegraphics[width=0.48\textwidth]{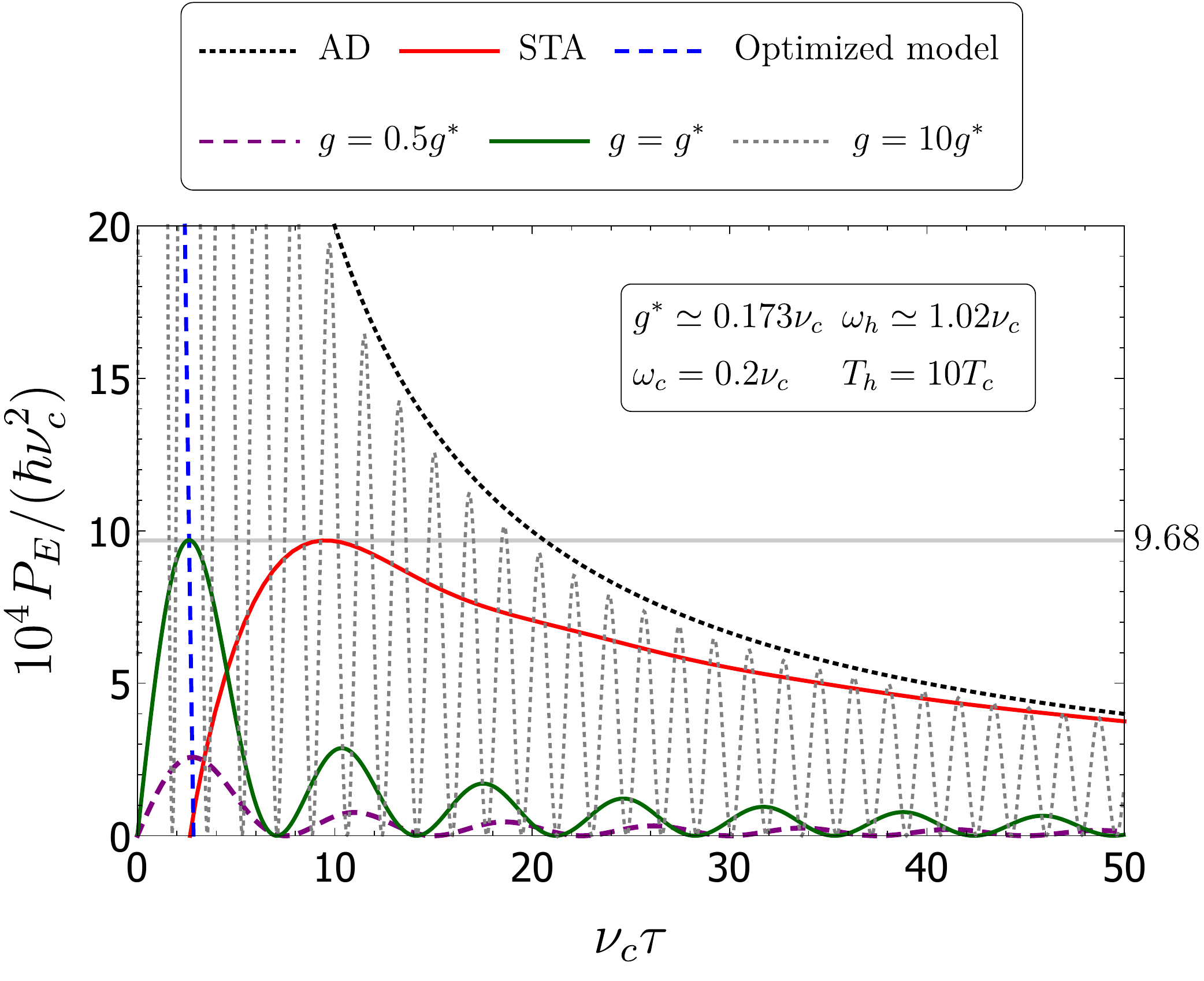}
	\caption{Power $P_E$ as a function of the total cycle time, which is equal to the collision time $\tau$ in our model without the mediator and to twice the driving time in Ref.~\cite{Cakmak2019}  ($t_w$ is assumed to be negligible compared to $\tau$), for the parameters given in the inset and in the legend, in the case of qubits. The continuous red line is the power obtained in the Otto cycle with STA. The dotted black line is the power in an ideal Otto cycle without friction, i.e., an Otto cycle in which one makes the adiabatic approximation (AD line). The dashed purple line, the continuous green line, and the dotted grey line  (all three oscillating) refer to the power obtained with our model without the mediator, for various fixed coupling strengths $g$. The dashed blue line is the ensemble of the peaks of the oscillating lines for all the values of $g$ (see text). The maximum power of the Otto cycle with STA  ($\simeq 9.68 \times 10^{-4}\hbar\nu_c^2$, indicated in the plot by a continuous horizontal line) can be obtained in our model for $g \simeq 0.173 \nu_c$.}
	\label{fig:ComparisonCakmak}
\end{figure}

We start by comparing a qubit-powered engine fixing the same energy separations and temperatures of Ref.~\cite{Cakmak2019}. Fig.~\ref{fig:ComparisonCakmak} shows various curves of power as functions of the total cycle time, which we denote with $\tau$ since, being the waiting times assumed to be negligible for all the curves (as in Ref.~\cite{Cakmak2019}), it coincides with the collision time $\tau$. With respect to the model of Ref.~\cite{Cakmak2019}, $\tau$ corresponds to twice the driving time. The continuous red line shows the power of the non-adiabatic quantum Otto engine improved by the STA developed in Ref.~\cite{Cakmak2019}. Instead, the dotted black line is the power that one would obtain if the adiabatic approximation worked even for finite times (AD line). The dashed purple line, the continuous green line, and the dotted grey line (all three oscillating) give, respectively, the power $P_E$ of our cycle without mediator [cf. Eq.~\eqref{eq:PowerFormula}] as a function of $\tau$ for different values of the coupling $g$. The quantity $g^*$ is obtained by searching for the value of $g$ such that the peak of power in our model is roughly equal to that of the STA cycle. Finally, the dashed blue line represents the collection of the maximum values of the oscillating lines for all coupling strengths. In other words, it is the parametric curve $(\tau^*(g),P_{E,\textrm{max}}(g))$, reported in the plot with appropriate units, where $\tau^*(g)$ is the optimal collision time given the coupling $g$, satisfying $k \tau^*(g)=y*$ [see Sec.~\ref{subsec:MaximizationFixedFrequencies}]. This plot shows that in our model one does not need to have $g \gg \omega_c, \omega_h$ to make the maximum power comparable with that of the quantum Otto cycle. Indeed, in this example, when $g \simeq 0.173 \nu_c$, our cycle gives roughly the same peak of power output of the Otto cycle while maintaining the efficiency of an ideal Otto cycle. Moreover, our design does not require the control of time-dependent Hamiltonians, needed to implement Otto cycles with STA.

Figure~\ref{fig:ComparisonCakmak} also shows that for large values of $g$, the local peaks of the power approach the values obtainable in the adiabatic limit. This stems from the fact that the power of the ideal Otto Cycle without friction (i.e., an Otto cycle which we treat as adiabatic even for finite times) is equal to that of Eq.~\eqref{eq:PowerFormula} with $V=1/\tau$ \cite{Cakmak2019} and that, when $k \tau = \pi/2+ m\pi$, where $m$ is a natural number, $V=g^2/(k^2 \tau)$ [cf. Eq.~\eqref{eq:Vterm}, recalling that here $t_w=0$].

\begin{figure}
	\centering
	\includegraphics[width=0.48\textwidth]{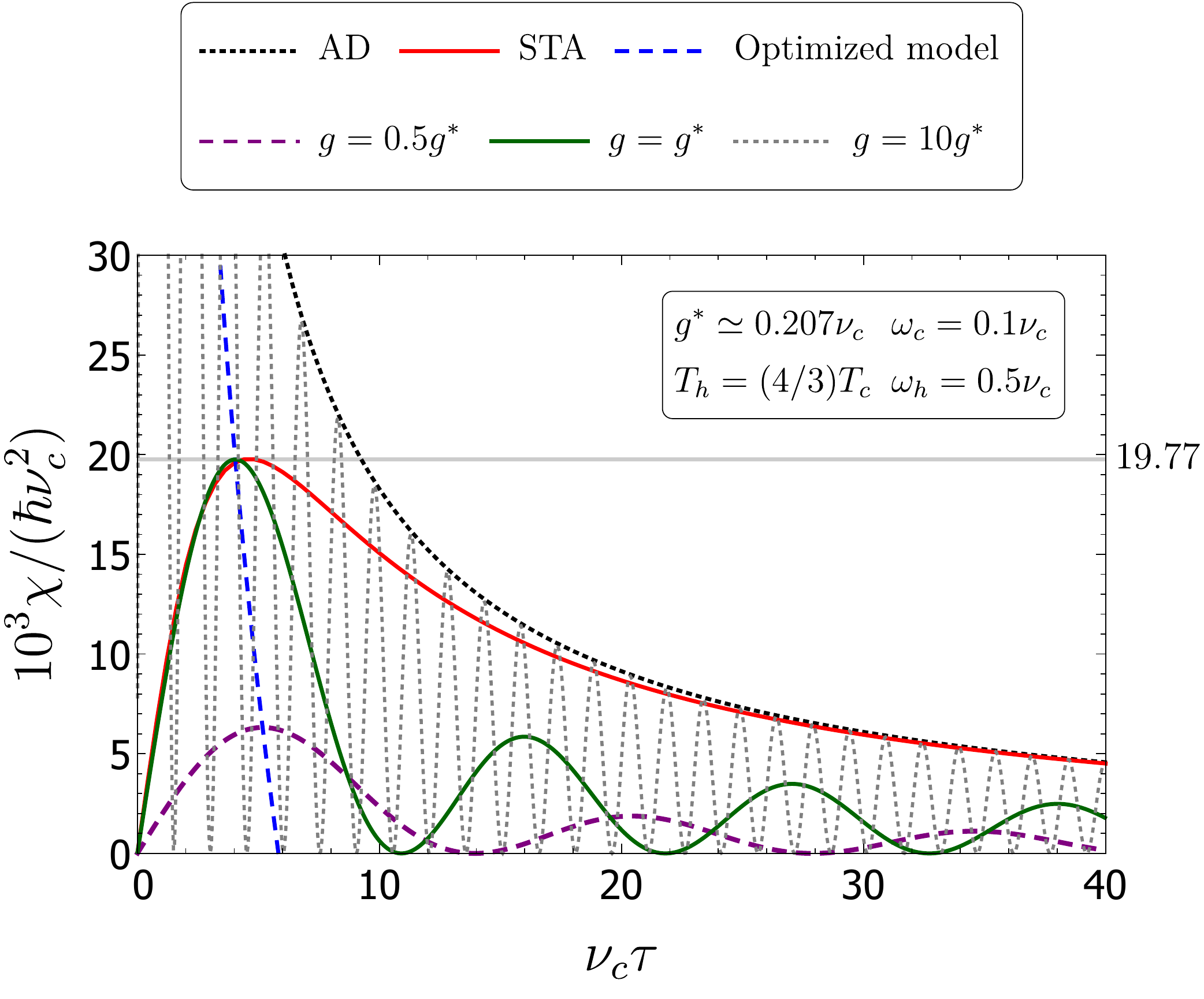}
	\caption{ Figure of merit $\chi$ as a function of the total cycle time, which is equal to the collision time $\tau$ in our model without the mediator and to twice the driving time in Ref.~\cite{Abah2020} ($t_w$ is assumed to be negligible compared to $\tau$), for the parameters given in the inset and in the legend, in the case of quantum harmonic oscillators. The continuous red line is the figure of merit obtained in the Otto cycle with STA. The dotted black line is the figure of merit in an ideal Otto cycle without friction, i.e., an Otto cycle in which one makes the adiabatic approximation (AD line). The dashed purple line, the continuous green line, and the dotted grey line  (all three oscillating) refer to the figure of merit obtained with our model without the mediator, for various fixed coupling strengths $g$. The dashed blue line is the ensemble of the peaks of the oscillating lines for all the values of $g$ [see part of the text concerning the same line in Fig.~\ref{fig:ComparisonCakmak}]. The maximum figure of merit of the Otto cycle with STA ($ \simeq 19.77\times 10^{-3}\hbar\nu_c^2$, indicated in the plot by a continuous horizontal line) can be obtained in our model for $g \simeq 0.207  \nu_c$.}
	\label{fig:ComparisonAbah}
\end{figure}

Fig.~\ref{fig:ComparisonAbah} shows the same kind of comparison in the refrigerator case but for harmonic oscillators, as in Ref.~\cite{Abah2020}. Instead of the power, we have plotted the figure of merit $\chi = \epsilon_R P_R$, where $\epsilon_R$ is the COP and $P_R= Q_c/{\tau}$ is the cooling power. Analogously to the previous comparison, the condition $g \gg \omega_c,\omega_h$ is not needed to obtain the same peak value of $\chi$ obtained in the quantum Otto cycle. Note that for $g \geq \sqrt{\omega_c \omega_h}\simeq 0.224 \nu_c$ the interaction Hamiltonian between two harmonic oscillators  leads to an instability (see Appendix~\ref{APPsec:CouplingLimit}).

\section{\label{sec:Conclusions}Conclusions}

We have examined two different versions of a two-stroke quantum thermal machine. In both versions, two collections of identical systems with evenly spaced non-variable energy levels can be put in contact, respectively, with a cold and a hot thermal bath. Because of the evenly spaced energy levels, we can characterize each system through a single frequency. In the first version, a system of a collection interacts with a system of the other one, and then they thermalize. In the second one, we have added a mediator system that interacts alternately with one or more systems of each collection.

Assuming the interaction Hamiltonian to conserve the number of excitations during the interaction, we have shown that the efficiency depends only on the frequencies of the two collections in both versions of the cycle.

In Sec.~\ref{sec:ModelNoMediator}, we have analyzed the problem of power maximization in the cycle without the mediator, focusing on the high impact of the waiting time on the optimal collision time and the optimal frequencies. When maximizing over the frequencies, we have shown that it is possible to exceed the Curzon-Ahlborn efficiency when using qubits while it is not possible with harmonic oscillators.

In Sec.~\ref{sec:CycleMediator}, we have added the mediator and we have shown that its addition can bring performance advantages when the waiting time of the corresponding cycle without the mediator is of the same order of the collision time. In most other cases, the cycle without the mediator performs better. However, there could be experimental platforms where it is easier to implement the cycle with the mediator.

Finally, in Sec.~\ref{sec:ComparisonOttoCycle}, we have compared our cycle without the mediator with two examples of Otto cycles enhanced by shortcuts to adiabaticity. The comparison has shown that one does not need high coupling ($g \gg \omega_c, \omega_h$) to obtain a power output comparable to that of the Otto cycle with shortcuts to adiabaticity at maximum power.

\section*{Acknowledgments}

N.P. acknowledges  the \enquote{Universit\'e de Franche-Comt\'e} for financial support throughout the mobility grant  \enquote{MOBILIT\'E INTERNATIONALE DES DOCTORANTS 2019} for his visit at the Queen's University Belfast (QUB) and thanks the QTeQ group and the CTAMOP unit at QUB for their kind ospitality on this occasion. 
G.D.C. acknowledges support from the UK EPSRC grants EP/S02994X/1 and EP/T026715/1 and discussions with Obinna  Abah, Adam Hewgill, Heather Leitch, and Mauro Paternostro.
 B.B. acknowledges  support by the French ``Investissements d'Avenir'' program, Project ISITE-BFC (Contract No. ANR-15-IDEX-03). N.P. and B.B. acknowledge Daniel Braun for private communication and interesting discussions on the topics of this paper.

\appendix

\section{\label{APPsec:CalculationCollision} Dynamics of a collision}

In this appendix, we describe in detail the dynamics of a collision between two systems governed by the Hamiltonian $H_T = H_c + H_h + H_I$, equal to
\begin{equation}
\label{APPeq:HamiltonianTotal}
H_T = \hbar \prtq{\omega_c \nc + \omega_h \nh + g\prt{\acd \ah + \ac \ahd}}.
\end{equation}
To this aim we compute, in the Heisenberg picture, the time evolution of the number operator for a system $S_h$, which we denote as $(\nh)_H (t)$. In order to do this, we first calculate the commutators of the following operators
\begin{align}
&\nh = \ahd \ah,\
&\nc = \acd \ac, \nonumber \\
&A_+ = \ah \acd + \ahd \ac,\
&A_- = \ah \acd - \ahd \ac,
\end{align}
obtaining
\begin{align}
\comm{\nh}{A_+} &= -A_-,\qquad 	\comm{\nc}{A_+}=  A_-, \nonumber \\
\comm{\nh}{A_-} &= - A_+,\qquad	\comm{\nc}{A_-}= A_+, \nonumber \\
\comm{A_+}{A_-} &= 2 \prtq{\nh \prt{1-d_c B_c} - \nc \prt{1 - d_h B_h}},
\end{align}
where $B_r = \dyad{N_r - 1}$. Notice that when ${d}_c = d_h = 0,2 $ (two-harmonic oscillator and two-qubit cases) one finds $\comm{A_+}{A_-} = 2 \prt{\nh - \nc}$, which gives rise to a closed algebra of commutators. In the following, we solve the dynamics analytically using this form of the commutator $\comm{A_+}{A_-}$. However, we will give some comments about the validity of the solutions in the general case. 
We observe that the commutator $\comm{A_+}{A_-}$ has not the above simplified form in the case of the Jaynes-Cumming model where a qubit interacts with an harmonic oscllator, so that the analytic solution of the dynamics presented in the remainder of this appendix cannot be applied to this case.

By using $\comm{A_+}{A_-} = 2 \prt{\nh - \nc}$ and by rewriting the total Hamiltonian of the system  as
\begin{equation}
H_T = \hbar \prt{\omega_c \nc + \omega_h \nh + g A_+},
\end{equation}
we obtain the following commutators
\begin{align}
	\comm{H_T}{\nh} /\hbar &= g  A_-, \quad
	\comm{H_T}{\nc} /\hbar = -g A_-, \nonumber \\
	\comm{H_T}{A_+} /\hbar &= \prt{\omega_c - \omega_h}A_-, \nonumber\\
	\comm{H_T}{A_-} /\hbar &= \prt{\omega_c - \omega_h}A_+ + 2 g \prt{\nh - \nc} .
\end{align}
Within this assumption, the operator $(\nh)_H (t)$ in the Heisenberg picture can be expressed by
\begin{equation}
\label{APPeq:OperatorHeisenbergForm}
\prt{\nh}_H (t) = f_h (t) \nh + f_c (t) \nc + f_+ (t) A_+ + f_- (t) A_-,
\end{equation}
where $f_h (0) = 1$ and $f_c (0) = f_+ (0) = f_- (0) = 0$.
The evolution of a generic operator $O$ in the Heisenberg picture is given by the Heisenberg formula
\begin{equation}
\label{APPeq:HeisenbergEquation}
\dot{O}_H (t) = (i/\hbar) \comm{H_T}{O_H (t)},
\end{equation}
where the corresponding operator $O$ is a time-independent operator in the Schr\"{o}dinger picture.
By inserting Eq.~\eqref{APPeq:OperatorHeisenbergForm} into Eq.~\eqref{APPeq:HeisenbergEquation} we obtain the following system of differential equations
\begin{equation}
\begin{cases}
\dot{f}_h(t) &=2 i g f_-(t),\\
\dot{f}_c(t) &= -2 i g f_-(t),\\
\dot{f}_+(t)&= - 2 i \delta f_-(t),\\
\dot{f}_-(t) &= - 2 i \delta f_+ (t) + i g \prtq{f_h (t)- f_c(t)},
\end{cases}
\end{equation} 
which together with the boundary conditions given above have the solutions
\begin{equation}
\begin{cases}
f_h(t) &= \frac{2 \delta^2 + g^2 \prtq{1+\cos\prt{2 k t}}}{2 k^2},\\
f_c(t) &= \frac{g^2 \prtq{1-\cos\prt{2 k t}}}{2 k^2},\\
f_+ (t)&= \frac{g \delta \prtq{1-\cos\prt{2 k t}}}{2 k^2},\\
f_-(t) &= i \frac{g \sin\prt{2 k t}}{2 k},
\end{cases}
\end{equation}
where we recall that $\delta = \prt{\omega_h - \omega_c}/2$ and $k = \sqrt{g^2 + \delta^2}$.

Now, we can calculate the value of $ \ev{(\nh)_H (t=\tau)} = \ev{\nh}_{\tau}$ after the interaction time $\tau$.
In particular, we focus on the case treated throughout the paper in which both systems are initially in  a thermal state. It follows that at the beginning of the collision   $\ev{A_+}_{0} = \ev{A_-}_{0} = 0$ and we get
\begin{equation}
\ev{\nh}_{\tau} = f_h (\tau) \ev{\nh}_\thrm + f_c (\tau) \ev{\nc}_\thrm.
\end{equation}
Noticing that $f_c(\tau) =1 - f_h (\tau)$ and denoting $A \equiv f_h(\tau)$, we can write
\begin{equation}
\label{APPeq:CollisionEvolution}
\ev{\nh}_{\tau} = \ev{\nc}_\thrm + \prt{\ev{\nh}_\thrm - \ev{\nc}_\thrm}A.
\end{equation}

Let us now comment on the approximate validity of the above solution when   $\comm{A_+}{A_-} \neq 2 \prt{\nh - \nc}$ (i.e., when one of the two conditions $d_c = d_h = 0$ or $d_c = d_h = 2 $ is not satisfied). 
In fact, we expect that when a system with evenly spaced energy levels is big enough and its higher energy levels are approximately empty during the dynamics, this system should be a good approximation of a harmonic oscillator. Then, if both systems satisfy this requirement, we expect that the solution found above describes the dynamics of the operators quite well.
However, one situation in which the above dynamics is surely wrong is when $\delta=0$ and $\max(\ev{\nc}_\thrm,\ev{\nh}_\thrm) > \min(N_{c},N_{h})$ because at some point we would have a system with more excitations than levels, which is absurd.
In general, we expect that the dynamics of Eq.~\eqref{APPeq:CollisionEvolution} is a good description if $\ev{B_r} \simeq 0$ during the whole evolution. 

\section{\label{APPsec:EfficienciesMediator}Functioning regimes and efficiency of the cycle with mediator}

In this Appendix, we analyze the efficiencies and energy fluxes of the cycle with the mediator described in Sec.~\ref{sec:CycleMediator}. This is done directly for the steady cycle.

After ${u}_r$ collisions with systems $S_r$, the difference in internal energy of system $S_{m}$ is given by the change of its average excitation number multiplied by the energy gap, i.e.,
\begin{equation}
\label{APPeq:MediatorInternalEnergy}
\Delta U_r = \hbar\omega_{m} \prt{ \ev{\ns}_{{u_r}}  - \ev{\ns}_{{0_r}} },
\end{equation}
where $\ev{\ns}_{{u_r}}$ is the average number of excitations in system $S_{m}$ after ${u}_r$ collisions with systems $S_r$ while $\ev{\ns}_{{0_r}}$ is the same quantity before the first collision.
The internal energy change of systems $S_r$ can be regarded as heat because, after their collision, these systems will thermalize again by being in contact with one thermal bath.
By considering the heat positive when energy flows from the bath to the system and taking into account
the conservation of the total excitation number, heat is given by
\begin{equation}
\label{APPeq:HeatMediatorCycle}
Q_r = -\hbar\omega_r \sum_{i=1}^{{u}_r} \prt{\ev{\nr}_{{i_r}} - \ev{\nr}_\thrm}=\frac{\omega_r}{\omega_{m}} \Delta U_r,
\end{equation}
where $\ev{\nr}_{{i_r}}$ is the average excitation number of system $S_r$ after the collision with system $S_{m}${, which has} already done ${i_r}-1$ collisions{,} and ${\ev{\nr}_\thrm}$ is the same quantity in the thermal state at temperature $T_r${, $\rho^\thrm_r $}.
Finally, by exploiting the first law of thermodynamics, we find that the work done in ${u}_r$ collisions is given by
\begin{equation}
\label{APPeq:workStrokesMediator}
W_r = \Delta U_r - Q_r = \hbar \prt{\omega_{m} - \omega_r}\prt{\ev{\ns}_{{u_r}} - \ev{\ns}_{{0_r}}}.
\end{equation}

Then, by using $\ev{\ns}_{{0_c}}=\ev{\ns}_{{u_h}}$ and $\ev{\ns}_{{0_h}}=\ev{\ns}_{{u_c}}$, the total work in a cycle is given by
\begin{equation}
\label{APPeq:WorkPerCycle}
W = W_c + W_h =\hbar \prt{\omega_c - \omega_h}\prt{\ev{\ns}_{{u_h}} - \ev{\ns}_{{u_c}}}.
\end{equation}
Alternatively, since in a cycle $\Delta U_c + \Delta U_h = 0$,  we can write
\begin{equation}
\label{APPeq:WorkEquivalences}
{Q_c = - \frac{\omega_c}{\omega_h} Q_h, \quad W = - \prt{1 - \frac{\omega_c}{\omega_h}}Q_h,}
\end{equation}
from which it is easy to obtain the efficiencies of Table~\ref{table:MachineFrequencies}.

Regarding the frequency conditions, to obtain them we must also impose two conditions for the collisions $S_{m}$-$S_r$.
On the one hand, the system $S_{m}$ has to acquire excitations if it has less than $S_r$ and vice versa. On the other hand, the system which had less excitations cannot have, after the collision, more than the other one had. In this case, it holds  $\ev{\ns}_{{u_h}} > \ev{\ns}_{{u_c}} \iff \ev{\nh}_{{\thrm}} > \ev{\nc}_{{\thrm}}$. Then, if  $N_c = N_h$, the sign of Eq.~\eqref{APPeq:WorkPerCycle} is completely determined by the value of the ratios $\omega_c/T_c$ and $\omega_h/T_h$ since the number of excitations in the mediator system has to be between $\ev{\nc}_\thrm$ and $\ev{\nh}_\thrm$ once the steady cycle has been reached.

\section{\label{APPsec:PerfectSwap} Comparison with a perfect swap Hamiltonian}

In our setting, when the parameter $g$ is very high ($g\gg \delta$), the collision between two systems approximately results  in a swap operation.
One could wonder if a Hamiltonian that generates the swap operation outperforms the exchange one proposed in Sec.~\ref{sec:ModelNoMediator}.
In this Appendix, we investigate this question for the case of two qubits.

In order to find a swap Hamiltonian, we first write a swap operation.
Using the basis $\{\ket{1_c,1_h},\ket{1_c,0_h},\ket{0_c,1_h},\ket{0_c,0_h}\}$, where $\ket{1_r}$ is the excited state of qubit $r$ and $\ket{0_r}$ is the ground one,
we choose the following swap operator:
\begin{equation}
U_{\textrm{SWAP}} = \mqty(
1 & 0 & 0 & 0 \\
0 & 0 & 1 & 0 \\
0 & 1 & 0 & 0 \\
0 & 0 & 0 & 1 ).
\end{equation}
The Hamiltonian which generates this unitary evolution after a swap time $t=t_S$  is \begin{equation}
H_{\textrm{SWAP}} = \frac{i \hbar}{{t_S}} \ln(U_{\textrm{SWAP}}) = \frac{\pi \hbar}{2 t_S}\mqty(
0 & 0 & 0 & 0 \\
0 & -1 & 1 & 0 \\
0 & 1 & -1 & 0 \\
0 & 0 & 0 & 0 ),
\end{equation}
which has spectrum $\sigma(H_{\textrm{SWAP}}) = \{-\pi \hbar / t_S,0,0,0\}$ and commutes with the total number operator, thus implying the same efficiencies of the exchange Hamiltonian (for the same frequencies) even when not performing a complete swap. Notice that we assume that during the collision the free Hamiltonians of the colliding systems are suppressed and the dynamics is governed only by the Hamiltonian $H_{\rm SWAP}$.

After a collision lasting  $\tau_S$, the swap Hamiltonian leads to the new populations
\begin{align}
\ev{\nc}_{\tau_S} &= \ev{\nc}_\thrm \cos^2\left(\frac{\pi \tau_S}{2 t_S}\right) + \ev{\nh}_\thrm \sin^2 \left(\frac{\pi \tau_S}{2 t_S}\right), \nonumber \\
\ev{\nh}_{\tau_S} &= \ev{\nh}_\thrm \cos^2 \left(\frac{\pi \tau_S}{2 t_S}\right) + \ev{\nc}_\thrm \sin^2 \left(\frac{\pi \tau_S}{2 t_S}\right).
\end{align}
By inserting the above equations into Eqs.~\eqref{eq:WorkNM} and \eqref{eq:HeatDefinition} we get the same formulas of Eq.~\eqref{eq:PowerFormula} but with the term $V$ substituted by
\begin{equation}
V_{S} = \frac{1}{\tau_S + t_w}\sin^2 \left(\frac{\pi \tau_S}{2 t_S}\right).
\end{equation}

To make a fair comparison we choose the parameter $t_S$ of the swap Hamiltonian so that the difference between maximum and minimum eigenvalues are equal to the same quantity in the case of the Hamiltonian of Eq.~\eqref{APPeq:HamiltonianTotal}, which has matrix form 
\begin{equation}
H_T =\hbar \mqty(
2 \bar{\omega} & 0 & 0 & 0 \\
0 & \omega_c & g & 0 \\
0 & g & \omega_h & 0 \\
0 & 0 & 0 & 0 ),
\end{equation}
and the spectrum of which is $\sigma (H_T) = \{2 \hbar\bar{\omega}, \hbar(\bar{\omega}+ k),\hbar(\bar{\omega}- k),0\}$,
where we recall that $\bar{\omega} = (\omega_c + \omega_h)/2$,  $k = \sqrt{\delta^2 + g^2}$ and $\delta = \prt{\omega_h - \omega_c}/2$.
The difference between maximum and minimum eigenvalues for the swap Hamiltonian is just $\pi \hbar/{t_S}$ while for our Hamiltonian is $2 \hbar\bar{\omega}$ if $g{\leq} \sqrt{\omega_c \omega_h}$ and $2 \hbar k$ if $g \geq \sqrt{\omega_c \omega_h}$.
Then, for the two cases, we have ${t_S}=\pi/(2 \bar{\omega})$ and ${t_S}=\pi/(2 k)$, respectively.
In both cases, if the collision time of the swap Hamiltonian can be optimized, the swap interaction outperforms the exchange one since, for $\abs{\delta}>0 $, 
if $t_S = \pi/(2\bar{\omega})$, which occurs when $\bar{\omega} > k$,
\begin{multline}
\max_{\tau} V = \frac{g^2}{k^2} \frac{\sin^2\prt{k \tau^*}}{\tau^* + t_w} <
\frac{g^2}{k^2} \frac{\sin^2\prt{k \tau^*}}{(k/\tilde{\omega})\tau^* + t_w}  \\
= \frac{g^2}{k^2} \frac{\sin^2\prt{\bar{\omega} \tau'}}{\tau'+ t_w}
\leq 
\frac{g^2}{k^2} \max_{\tau_S} \frac{\sin^2 \prt{\bar{\omega} \tau_S}}{\tau_S + t_w} <\max_{\tau_S} V_S,
\end{multline}
where $\tau' = (k/\bar{\omega}) \tau^*$, while, if $t_S=\pi/(2k)$,
\begin{equation}
\max_{\tau_S} V_S = \max_{\tau_S} \frac{\sin^2\prt{k \tau_S}}{\tau_S + t_w}  > \frac{g^2}{k^2} \max_{\tau} \frac{ \sin^2\prt{k \tau}}{\tau + t_w} = \max_{\tau} V.
\end{equation}

With respect to a perfect swap situation with switching time $\tau_S= t_S$, the time-optimized exchange Hamiltonian can give a higher power output. We show this in the case $t_w = 0$. 
  In the case $t_S = \pi/(2 \bar{\omega})$, one gets 
 that $\Vm > V_S (\tau_S = t_S)$ implies
\begin{equation}
g > \frac{\sqrt{2\bar{\omega} \prt{\bar{\omega} + \sqrt{\bar{\omega}^2 + \alpha^2 \pi^2 \delta^2}}}}{\alpha \pi},
\end{equation}
where $\alpha$ is the maximum value of $\sin^2(x)/x$, i.e., $\alpha \simeq 0.7246$.
Numerically, one can see that the condition $g \leq \sqrt{\omega_c \omega_h}$ can be satisfied together with the above one only when $\omega_h/\omega_c$ roughly lies in the interval $[0.4832,2.0697]$.
Regarding the ratio $\Vm/V_S(\tau_S = t_S)$, recalling that $t_S = \pi/(2 \bar{\omega})$ for $g \leq \sqrt{\omega_c \omega_h}$, one gets
\begin{equation}
\frac{\Vm}{V_S \prt{\tau_S = t_S}} = \frac{\alpha \pi}{2} \frac{g^2}{k \bar{\omega}} \leq \alpha \pi \frac{\omega_c \omega_h}{2 \bar{\omega}^2}.
\end{equation}

Regarding the case $t_S = \pi/(2 k)$, one gets 
 \begin{equation}
\Vm > V_S \prt{\tau_S = t_S} 
\iff 
g > \sqrt{\frac{2}{\alpha \pi - 2}} \abs{\delta} \simeq 2.6898 \abs{\delta},
\end{equation}
and \begin{equation}
\frac{\Vm}{V_S \prt{\tau_S = t_S}} = \frac{\alpha \pi}{2} \frac{g^2}{k^2} \simeq 1.1382 \frac{g^2}{k^2}.
\end{equation}

\section{\label{APPsec:PowerHarmonicOscillators}The power of a harmonic oscillator couple is not frequency bounded}

In this Appendix, we show, for the case of harmonic oscillators, that the power increases monotonically by decreasing the frequencies while keeping fixed their ratio and the temperatures.
In order to do this, we start by writing $\omega_h = \omega_c/(1- \eta_E)$ and $T_h = T_c / (1 - \eta_C)$ so that we can write the maximum power for the engine for a given $\omega_c$ as 
\begin{equation}
\tilde{P}_E = k_B T_c \frac{\eta_E}{1-\eta_E} f\prt{x} \Vm,
\end{equation}
where 
\begin{equation}
f\prt{x} = x \prtq{\coth\prt{l x}-\coth\prt{x}},
\end{equation}
$x= \hbar \omega_c / (2 k_B T_c)$, and $l = (1-\eta_C)/(1-{\eta_E})$.

In order to show that the maximum power  is obtained in the limit $x\rightarrow 0$, we prove that $\partial_x \tilde{P}_E <0\ \forall \ x>0$.
The derivative of $\tilde{P}_E$ takes the form
\begin{equation}\label{APPeq:PowerDerivative}
\partial_x \tilde{P}_E = k_B T_c \frac{\eta_E}{1-\eta_E} \prtq{\prt{\partial_x f\prt{x}} \Vm + f\prt{x} \partial_x \Vm}.
\end{equation}
We start by analyzing the term
\begin{equation}
\partial_x f(x) = \coth\prt{l x}-\coth\prt{x} + x \prtq{\csch^2\prt{x} - l \csch^2\prt{l x}},
\end{equation}
which can be shown to be negative for every $x>0$.
Indeed, we can rewrite the above function as
\begin{equation}
\partial_x f\prt{x} = b\prt{lx} - b\prt{x}, \quad
b\prt{x} = \frac{\sinh\prt{x} \cosh\prt{x} - x}{\sinh^2\prt{x}}.
\end{equation}
Since $0 \leq l \leq 1$ we can find the sign of $\partial_x f(x)$ by understanding the behavior of $b(x)$. If $b(x)$ is always increasing, it follows that $\partial_x f(x)$ is always negative.
Then, we calculate the derivative of $b(x)$:
\begin{equation}
\partial_x b (x) = \frac{2}{\sinh^2\prt{x}} \prtq{x \coth\prt{x} - 1},
\end{equation}
which is always positive because
\begin{equation}
x \coth\prt{x} - 1 > 0, \quad \forall \ x>0.
\end{equation}
Thus, since $\Vm > 0$, we have shown that the first term in the square brackets of Eq.~\eqref{APPeq:PowerDerivative} is negative. 
	
Concerning the second term in the square brackets of Eq.~\eqref{APPeq:PowerDerivative}, since $f(x) > 0  \ \forall \ x>0$, we could show that it is negative by proving that $\partial_x \Vm < 0 \ \forall \ x>0$.
Writing $\omega_h = \omega_c / (1-\eta_E)$, the quantity $k$ can be cast in the form
\begin{equation}\label{APPeq:ktermOmegac}
k= \sqrt{g^2 + \prt{\frac{\omega_c}{2} \frac{\eta_E}{1-\eta_E}}^2},
\end{equation}
which makes evident that $k$ increases when $\omega_c$ does, i.e., $\partial_x k > 0 \ \forall x>0$.
Then, we can just study [cf. Eq.~\eqref{eq:Vterm}]
\begin{equations}
\partial_k \Vm = \partial_k \prt{\frac{g^2}{k} \frac{\sin^2 \left(z \right)}{z+k t_w}},
\end{equations}
where $z= k \tau^*$ is the value for which $\Vm$ is maximized for given $k$ and $t_w$.
For $t_w = 0$, $z=y^*$ [cf. Sec.~\ref{subsec:MaximizationFixedFrequencies}] so that $\partial_k z = 0$ and the disequality $\partial_k \Vm < 0$  reduces to $- \sin^2(y^*) < 0$ which is, indeed, true.
For $t_w > 0$ we can write
\begin{equations}
\partial_k \Vm = - \frac{g^2}{k^2} \frac{\sin^2\prt{z}}{z+k t_w} + \frac{g^2}{k} \partial_k \prt{\frac{\sin^2\prt{z}}{z+k t_w}},
\end{equations}
where the first term is negative since $k>0$ and $y^* < z < \pi/2$.
The second term can be rewritten as
\begin{equation}
\frac{g^2 t_w}{k} \partial_{\prt{k t_w}} \prt{\frac{\sin^2\prt{z}}{z+k t_w}},
\end{equation}
and one can numerically check that is negative for $k t_w >0$.
Then, $\partial_k \Vm < 0$ follows and, being $\partial_x \tilde{P}_E$ composed by the sum of two negative quantities multiplied by a positive one, it also follows that $\partial_x \tilde{P}_E < 0 \ \forall \ x > 0$.

\section{\label{APPsec:PowerMaximization}  Maximization of power in the cycle with the mediator}

In this Appendix, we deal with the problem of maximizing the power of the cycle with the mediator, when the frequencies of the collections, the temperatures, and the couplings are fixed. To this aim,
we must first find the stationary values of the average excitation number of system $S_{m}$ in the steady cycle.

We already know, from Appendix~\ref{APPsec:CalculationCollision} that the result of each collision between system $S_{m}$ and system $S_r$ is described by the analogous of Eq.~\eqref{APPeq:CollisionEvolution} if $\ev{A_-}_{0} = \ev{A_+}_{0} = 0$ at the beginning of the collision under examination (we recall that this solution is exact only for the cases of qubits and harmonic oscillators).

We notice that this condition is always satisfied since systems $S_r$ are in the thermal state before their collision. Then, one can easily prove by induction that for ${u}_r$ collisions of the same duration the evolution is given by
\begin{equation}
\label{APPeq:MediatorCollision}
\ev{\ns}_{{u_r}} = \ev{n_r}_\thrm + \prt{\ev{\ns}_{{0_r}} - \ev{n_r}_\thrm}A_r^{{u_r}},
\end{equation}
where $A_r$ is defined below Eq.~\eqref{eq:VtermMediator}.
If the duration time of the collisions can vary, the term $A_r^{{u}_r}$ becomes
\begin{equation}
A_r^{u_r} \longrightarrow \prod_{i=1}^{u_r} A_r \prt{\tau_{r, i}}{,}
\end{equation}
where $\tau_{r,i}$ denotes the duration of the $i$-th collision.

Turning back to the case of equal-time collisions, we can find the steady cycle values of $\ev{\ns}_{{u_h}}$ and $\ev{\ns}_{{u_c}}$ by solving the following  simultaneous equations:
\begin{align}
	\ev{\ns}_{{u_h}} &= \ev{\nh}_\thrm
	 + \prt{\ev{\ns}_{{u_c}} - \ev{\nh}_\thrm}A_h^{{u_h}}, \nonumber \\
	\ev{\ns}_{{u_c}} &= \ev{\nc}_\thrm + \prt{\ev{\ns}_{{u_h}} - \ev{\nc}_\thrm}A_c^{{u_c}},
\end{align}
which lead to Eq.~\eqref{eq:VtermMediator} using Eq.~\eqref{APPeq:WorkPerCycle}.

Even if we cannot provide an analytical proof that one collision is the best option, performance-wise, we can provide an argument for why we strongly believe so.
In the cycle with the mediator, the work is connected to the change of internal energy of system $S_{m}$ [cf. Eqs.~\eqref{APPeq:HeatMediatorCycle} and \eqref{APPeq:workStrokesMediator}].
Then, instead of showing that $u_c = u_h = 1$ is optimal for maximizing Eq.~\eqref{eq:VtermMediator} we show that the maximization of $\abs{\Delta U_r}/(u_r \tau_r)$ is obtained for $u_r = 1$.
From Eqs.~\eqref{APPeq:MediatorInternalEnergy} and \eqref{APPeq:MediatorCollision} we get 
\begin{multline}
\frac{{\abs{\Delta U_r (1)}}}{\tau_r} - \frac{{\abs{\Delta U_r ({u}_r)}}}{{u}_r \tau_r} = \hbar \omega_{m} \left|\ev{\nr}_{{\thrm}}-\ev{\ns}_{{0_r}}\right|  \\ \times \prt{\frac{1 - A_r}{\tau_r} - \frac{1 - A_r^{{u}_r}}{{u}_r \tau_r}}{,}
\end{multline}
where $u_r \geq 2$ and $\Delta U_r ({u}_r)$ indicates $\Delta U_r$ after $u_r$ collisions.
If $A_r = 0$, the above quantity is positive, while it is zero for $A_r = 1$.
By deriving the above quantity with respect to $A_r$, we can see that the derivative  is always non-positive for $A_r \in [0,1]$ [this is the interval where $A_r$ is confined, cf. Eq.~\eqref{eq:OneCollision} and the comment below Eq.~\eqref{eq:VtermMediator}].
This means that when all the other parameters are fixed, the energy exchange rate with one collision is maximized for ${u}_r = 1$.
We remark that the above argument refers only to one stroke of the cycle, and then it is not a proof that the maximization of the entire cycle involving a proper maximization of the term $V_m$ [Eq.~\eqref{eq:VtermMediator}] is obtained for ${u}_c = u_h = 1$.
However, Fig.~\ref{fig:graphTimesWithMediator} and other numerical simulations suggest that this is indeed the case.

 We finally remark, in the case of qubits and harmonic oscillators (with systems of the same kind for both the mediator and the collection), that in the limit case of a great number of collisions between the mediator and systems of a collection, the mediator steady state has the same average number of excitations of the thermal state of the collection systems [cf. Eq.~\eqref{APPeq:MediatorCollision}]. It is also possible to show that these states have the same populations level by level, as already predicted for harmonic oscillators in Ref.~\cite{Grimmer2018}.

\section{\label{APPsec:CouplingLimit}Coupling limit for quantum harmonic oscillators}

Here, we discuss the physical limit with respect to the coupling of the total Hamiltonian in the case of harmonic oscillators. 

When considering the interaction of two harmonic oscillators, the term $g$ cannot be too high because, otherwise, one of the so-called normal frequencies becomes negative and this translates into a total Hamiltonian not bounded from below~\cite{Estes1968,BookSerafini2017}.
Referring to the Gaussian formalism~\cite{BookSerafini2017}, this translates into the requirement of positive definiteness for the \enquote{Hamiltonian matrix}.
Assuming the exchange interaction of Eq.~\eqref{eq:InteractionHamiltonian}, the total Hamiltonian is given by Eq.~\eqref{APPeq:HamiltonianTotal} and its  \enquote{Hamiltonian matrix} (with respect to the quadrature operators, see Ref.~\cite{BookSerafini2017}) reads 
\begin{equation}
\hbar \mqty(
\omega_h & 0 & g & 0\\
0 & \omega_h & 0 & g\\
g & 0 & \omega_c & 0\\
0 & g & 0 & \omega_c\\
){.}
\end{equation}
The doubly degenerate eigenvalues of the above matrix are 
\begin{equation}
\frac{\lambda_\pm}{\hbar}  = \frac{\omega_h + \omega_c}{2} \pm \sqrt{\delta^2 + g^2},
\end{equation}
thus leading to the condition $g < \sqrt{\omega_c \omega_h}$ to avoid non-positive  normal frequencies.
However, if the two systems are not real harmonic oscillators but many-level ones, the highest levels of which are practically unoccupied during the whole dynamics, they are very well approximated by harmonic oscillators and this problem can be avoided.


%

\end{document}